\documentclass{emulateapj}
\usepackage{graphicx}
\usepackage{color}
\usepackage{amsmath}
\usepackage{amssymb}
\usepackage{hyperref}
\usepackage{epstopdf}
\usepackage{upgreek}
\usepackage{xcolor}
\hypersetup{
    colorlinks,
    linkcolor={red!80!black},
    citecolor={blue!80!black},
    urlcolor={blue!80!black}
}

\usepackage{xcolor}

\def\wig#1{\mathrel{\hbox{\hbox to 0pt{%
          \lower.5ex\hbox{$\sim$}\hss}\raise.4ex\hbox{$#1$}}}}

\shorttitle{Inferring the Eccentricity Distribution of Short-Period Planet Candidates}
\shortauthors{Shabram, et al.}

\newcommand{\cp}{\citep}
\newcommand{\ct}{\citet}
\providecommand{\e}[1]{\ensuremath{\times10^{#1}}}

\newcommand{\changes}[1]{}

\slugcomment{Accepted for publication in ApJ}

\begin{document}

\title{The Eccentricity Distribution of Short-Period Planet Candidates Detected by \textit{Kepler} in Occultation}

\author{Megan Shabram\altaffilmark{1}, Brice-Olivier Demory\altaffilmark{2}, Jessi Cisewski\altaffilmark{3}, Eric B. Ford\altaffilmark{1}, Leslie Rogers\altaffilmark{4}} 

\altaffiltext{1}{Department of Astronomy and Astrophysics, Eberly College of Science, The Pennsylvania State University, 525 Davey Lab, University Park, PA 16802}
\altaffiltext{2}{Cavendish Laboratory, University of Cambridge, JJ Thomson Avenue, Cambridge, CB30HE UK}
\altaffiltext{3}{Department of Statistics, Carnegie Mellon University, Pittsburgh, PA 15214}
\altaffiltext{4}{California Institute of Technology, MC249-17, 1200 East California Boulevard, Pasadena, CA 91125, USA}

\begin{abstract} 

We characterize the eccentricity distribution of a sample of $\sim50$ short-period planet candidates using transit and occultation measurements from NASA's \textit{Kepler} Mission. 
First, we evaluate the sensitivity of our hierarchical Bayesian modeling and test its robustness to model misspecification using simulated data.
When analyzing actual data assuming a Rayleigh distribution for eccentricity, we find that the posterior mode for the dispersion parameter is $\sigma=0.081 \pm^{0.014}_{0.003}$.  
We find that a two-component Gaussian mixture model for $e \cos \omega$ and $e \sin \omega$ provides a better model than either a Rayleigh or Beta distribution.  Based on our favored model, we find that $\sim90\%$ of planet candidates in our sample come from a population with an eccentricity distribution characterized by a small dispersion ($\sim0.01$), and $\sim10\%$ come from a population with a larger dispersion ($\sim0.22$).  
Finally, we investigate how the eccentricity distribution correlates with selected planet and host star parameters.
We find evidence that suggests systems around higher metallicity stars and planet candidates with smaller radii come from a more complex eccentricity distribution.  

\end{abstract}

\keywords{methods: statistical; planets and satellites: dynamical evolution and stability, formation}

\section{Introduction}

The \textit{Kepler} mission has identified a sample of planet candidates detected both in transit and occultation, providing detailed orbital information, including orbital eccentricity, for a subset of systems with a wide variety of stellar host properties.
However, early works on the eccentricity distribution of all \textit{Kepler} objects of interest (KOIs), including those in this subset, are limited due to uncertainties in host star properties.
Recent studies have focused on applying Bayesian data analysis for robust error estimation \citep[e.g.,][]{Parviainen12}, and other studies have investigated the eccentricity distribution of planets discovered with the radial velocity technique and the role that tidal interactions play in shaping eccentricity distributions \citep{Wang11, Matsumura08, Hansen14}.
While some studies have attempted to constrain the eccentricity distribution of planets via transit durations identified by \textit{Kepler}, these studies have been limited by uncertainties in stellar densities \citep{Moorhead11, Kane12, Plavchan14, VanEylen15}.
\ct{Lucy12} used a Bayesian approach to explore the eccentricity distribution of eclipsing binaries. 
\ct{Kipping14b} explored biases in an eccentricity distribution using a Beta distribution prior, but little else has been done to explore the eccentricity distribution of exoplanets via similar methods and with the goal of quantifying population-level parameters. 
\ct{Hogg10} proposed using an hierarchical Bayesian (HB) model to constrain the eccentricity distribution of hot Jupiters, but applied their model to simulated radial velocity observations, only.  

Bayesian inference has made its way into exoplanet studies as computing facilities have evolved to accommodate the required calculations.
The application of HB modeling is highly relevant for studying the \textit{Kepler} planet sample  \citep[e.g.,][]{Demory14, Wolfgang14, Rogers14, DFM14}. 
This framework allows us to obtain population-level posterior distributions, such as the distribution function for planets, while accounting for measurement uncertainties and potentially, selection effects.
HB is particularly well suited for characterizing a population's eccentricity distribution largely because of its ability to accommodate samples where each measurement has a large measurement uncertainty. 

As a first step in studying the exoplanet population in general, we use HB modeling to investigate the eccentricity distribution of the subset \textit{Kepler} planet candidates that are detected in both transit and occultation, which provides measurements of projected eccentricity via transit duration ratios and phase offsets.
Even for this subset of planet candidates, individual eccentricity measurements often have large uncertainties.
Fittingly, HB is designed to account for individual measurement uncertainties.
Thus, we approach this problem from both sides: we will apply modern statistical methods that incorporate uncertainties into our eccentricity study (e.g., HB modeling), while also working with a subset of planet candidates with enough information to help bypass some of the uncertainty in their host star parameters.
The projected eccentricity measurements ($e \cos \omega$ and $e \sin \omega$) are presumed independent of the stellar host star density and radius, which mitigates the problem of uncertainties in stellar parameters.
Applying HB to the eccentricity distribution is a logical starting point while working to construct a comprehensive hierarchical model (i.e., a joint population distribution that includes planet parameters in addition to orbital eccentricity) in which, measurement uncertainties are naturally incorporated into the analysis. 

Furthermore, we can investigate various sub-populations of planets from the \textit{Kepler} sample and look for correlations of planet and host star properties within these subpopulations.  
In particular, we explore mixture models where the eccentricity distribution can be interpreted as a combination of two sub-populations.
With this analysis, more than one population in the eccentricity distribution could arise, for example, due to different formation mechanisms at work.   
Characterizing the eccentricity distribution in this way provides insight into postulated planet formation theories such as planet orbital migration and planet-planet scattering.  
In principle, these mechanisms could form two populations that make up the eccentricity distribution: one population that evolved via slow disk migration and another population that evolved via excitation of a large eccentricity (e.g., planet-planet scattering or secular perturbations) proceeded by tidal circularization.  
With this in mind, the population of planets that came from planet-planet scattering might have a larger dispersion as it would include planets with large eccentricities, while the population of planets that came from disk migration might have a smaller dispersion and contain fewer eccentric planets.   
These populations might also correlate with host star properties, which would allow for a framework to test physical models of the origin of each population thus shedding light onto planet formation. 

Here, we focus on inferring the eccentricity distribution of an interesting subset of planets using HB modeling applied to both simulated and real transit and occultation measurements from the \textit{Kepler} mission.   
This sample contains predominantly short-period planet candidates, most of which are likely to be hot Jupiters, identified by \textit{Kepler}.
We look for correlations between the eccentricity distribution and other properties, such as stellar effective temperature, planet radius, orbital period, and stellar metallicity to begin to synthesize a global understanding of planet formation.
This manuscript is organized as follows.
In \textsection 2, we describe  our observational data.
In \textsection 3, we describe the method behind the HB analysis calculations, and the priors selected for the study. 
In \textsection 4, we present the results of our HB analysis.
In \textsection 5, we investigate potential correlations between the eccentricity distribution and planet or host star properties.
In \textsection 6, we  summarize our results,
 and in \textsection 7, we discuss our conclusions, potential biases and future work. 

\section{Observations}

When a planet both transits and occults its host star, we are able to obtain detailed information about the planetary orbit, including information about the projected orbital eccentricity, $h=e \cos \omega$ and $k=e \sin \omega$. 
The relationship between orbital eccentricity and transit observables is outlined in \ct{Winn10}.  $h$ can be derived from
\begin{equation}\label{eq:phase_offset}
  \Delta t_c \approx \frac{P}{2} \left[ 1+\frac{4}{\pi} h \right],
\end{equation}
where $\Delta t_{c}$ is the time between the center of the transit and the center of the occultation and $P$ is the orbital period.  
$k$ can be derived from
\begin{equation}\label{eq:dur_ratio}
  \frac{ T_{\mathrm{occ}} }{ T_{\mathrm{tra}} } \approx \frac{1+k}{1-k},
\end{equation}
where $T_{occ}$ is the occultation duration and $T_{tra}$ is the transit duration. 
\changes{More precise expressions} for $h$ and $k$ are listed in \ct{Ragozzine09}, section 2.5 and elsewhere. 
\changes{The original full derivations of these expressions can be found in \citet{Sterne1940} and \citet{deKort1954}.}
When analyzing \textit{Kepler} observations in \textsection 5, we calculate the transit and occultation times and durations numerically using Keplerian orbits.

We have measured the offsets and durations of transits and occultations for a sample of planet candidates observed by \textit{Kepler}.
This study is based on quarters Q0 through Q12 \textit{Kepler} data \citep[see][for the Q0-Q10 data release]{Burke14}. In total, the datasets encompass about $\sim$1100 days of quasi-continuous photometric monitoring between May 2009 and March 2012. We retrieved the Q0-Q12 FITS files from MAST\footnote{http://archive.stsci.edu/kepler/} and extracted the SAP$\,_{FLUX}$, commonly known as ``calibrated light curves", long-cadence photometry \citep{Jenkins10} for each target. 
Using the calibrated data eliminates the potential instrumental corrections or cotrending basis vectors to introduce noise correlated on short timescales.

\subsection{Derivation of the planet physical and orbital properties}

We focus on a sample of planet candidates for which an occultation is detected. 
We address the biases that the selection effects introduce into our sample in \textsection 3.2 and \textsection 7.
We use the \textit{Kepler} planet candidate list to keep all planets larger than 8 Earth radii and with orbital periods less than 10 days. 
This initial selection of planet candidates was based on early planet candidate lists from NExSci. 
Note that most planets with a detectable occultation have very high SNR transits, so we do not expect that many additional planets would be found in the full Q0-Q17 datasets. 
The preliminary parameters were derived using the KIC stellar values, and updated later in our analysis. 
We employ a Markov Chain Monte Carlo (MCMC) framework to compute the posterior distribution of the system's orbital parameters using these initial values.
When performing MCMC analysis, we used an empirical main sequence mass-radius relationship \cp{Torres09} to derive more accurate planetary parameters. 
After the MCMC analysis was performed, some planet candidate radii changed to be outside the initial range stated above.
Our MCMC implementation (described in \citet{Gillon12}) uses the Gibbs sampler and the Metropolis-Hastings algorithm to estimate the posterior distribution function of all unknown parameters. Our nominal model is based on a star and a single transiting planet on a Keplerian orbit about their center of mass.

The input data provided to each MCMC run consist of the Q0-Q12 \textit{Kepler} photometry and the stellar parameters (effective temperature T$\,_{eff}$, metallicity $[Fe/H]$ and spectroscopic $\log g$) extracted from the \textit{Kepler} Input Catalog (KIC) \citep{Brown11}. We correct for the photometric dilution induced by neighboring stellar sources using a quarter-dependent dilution factor based on the dilution values presented in the literature and on the contamination values reported in the FITS files headers \citep{Bryson13}. 

We divide the total lightcurve in segments of duration $\sim$24 to 48 hrs. 
The smooth photometric variations due to stellar variability or instrumental systematic effects in each segment are fit with a time-dependent quadratic polynomial. 
Baseline polynomial coefficients are determined at each step of the MCMC for each lightcurve with the singular value decomposition method. 
The resulting coefficients are then used to correct the raw photometric lightcurves.
We assume a quadratic law for the limb-darkening (LD) and use $c_1=2u_1+u_2$ and $c_2=u_1-2u_2$ as jump parameters, where $u_1$ and $u_2$ are the quadratic coefficients \citep{Mandel02}.
We integrate over the 29.4 minute long cadence integration time when modeling long cadence light curves.

The MCMC has the following set of jump parameters (i.e., parameters that are not fixed in our model and are used as a basis for proposal steps): the planet/star flux ratio, the impact parameter $b$, the transit duration from first to fourth contact, the time of minimum light $T_0$, the orbital period, the occultation depth, the two LD combinations $c_1$ and $c_2$ and the two parameters $\sqrt{e}\cos\omega$ and $\sqrt{e}\sin\omega$.
At each step of the MCMC, the Keplerian model is constructed based on the $e$ and $\omega$ values derived from the $\sqrt{e}\cos\omega$ and $\sqrt{e}\sin\omega$ jump parameters.
A uniform prior distribution is assumed for all jump parameters except $c_1$ and $c_2$.
This corresponds to a prior that is uniform in $e \in [0,1)$ and $\omega \in [0,2\pi)$.
For the limb-darkening parameters, we assume normal priors which are centered on values of $c_1$ and $c_2$ that correspond to the values of $u_1$ and $u_2$ from the theoretical tables of \citet{Claret11} for the stellar parameters obtained from the KIC.
The standard deviation of the priors for $c_1$ and $c_2$ were set by the corresponding standard deviations propagated from $u_1$ and $u_2$'s uncertainties.
We run two Markov chains of 100,000 steps for each planet candidate. 
The mixing and convergence of the Markov chains are assessed using the Gelman-Rubin statistic criterion \citep{Gelman92}.
Results for $e \cos \omega$ and $e \sin \omega$ are shown in Table \ref{table1:data}. 

\subsection{Properties of Planet Candidates Analyzed}

When selecting the initial planet candidates that we perform MCMC fits for planet properties described above, we vet for eclipsing binaries (EBs) using the procedure outlined in \ct{Demory11}.
This leaves us with a sample of $85$ planet candidates for which we have calculated posteriors for their orbital and physical properties.
From this new list of planet candidates with updated properties from MCMC fitting, we do a second updated sweep for eclipsing binaries referring to \ct{Tenenbaum14} and \ct{Bryson13}, works that were published after our initial planet candidate list was developed.  
We also reference the \textit{Kepler} Eclipsing Binary catalog\footnote{http://exoplanetarchive.ipac.caltech.edu/docs/eclbin.html} for additional newly reported EBs.
From this procedure, we are able to exclude an additional $18$ planet candidates. 
We include KOI $1227$ in our sample of planet candidates as it appears in both the \textit{Kepler} eclipsing binary catalog with a period of $\sim4$ days, and in the \textit{Kepler} planet catalog as a potential planet with an $\sim2$ day period.  
After the vetting outlined above, we exclude an additional $17$ planet candidates for which the occultation signal-to-noise was low, resulting in very poor measurements of $h$ and $k$.  
This leaves us with $50$ planet candidates that have approximately Gaussian measurement uncertainties for $h$ and $k$ to use for our analysis of the eccentricity distribution in \textsection \ref{sec:results:occs}. 
Working in $h$ and $k$ space instead of eccentricity space greatly simplifies our HB model for the eccentricity distribution (see \textsection \ref{sec:method:HBM}), since the measurement uncertainties for $h$ and $k$ can be assumed to be roughly normally distributed.

The $50$ remaining planet candidates have radii estimates of $\sim1.9$ to $30$ Earth radii, with a median value of $10.6$ Earth radii, host star effective temperature of  $3948$ K to $8848$ K, with a median value of $5728$ K, orbital period of $1.03$ days to $20.13$ days with a median value of $4.24$ days, and host star metallicity of $-0.518$ to $0.440$ in [Fe/H] with a median value of $0.023$ [Fe/H].  
The $30$ Earth Radii planet candidate (KOI $1793$) is large, and an outlier for typical radii in our sample, but still makes it past our EB vetting procedure outlined above.  
These values came from the \textit{Kepler} Star Properties Catalog as reported at Exoplanet Archive updated December 2013 and revised February 2014 \cp{Buchhave12, Huber14}.  
The majority of the planets have stellar metallicity values obtained from photometry, and $10$ planets have spectroscopically derived stellar parameters.

\section{Method}

We aim to simulate and characterize the eccentricity distribution of a subset of the population of planet candidates in the \textit{Kepler} sample for which both transits and occultations have been observed. 
First, we describe a general HB model, before specializing it for our application of characterizing the eccentricity distribution in \textsection 3.2.
Next, we build and test a model using simulated data in order to determine the accuracy of our method and robustness to model misspecification in \textsection \ref{sec:results:ValidatingHBM}, then we apply our model to the real dataset in \textsection\ref{sec:results:occs}.

\subsection{The Hierarchical Bayesian Model} 
\label{sec:method:HBM}

Hierarchical Bayesian (HB) modeling is a powerful method to estimate population parameters by propagating the unique uncertainty from each measurement of the population constituents into the inference of the population parameters.  
An HB model requires an analysis model that parameterizes the functional form of the population distribution, $p(x_p | \boldsymbol{\phi})$, where $x_p$ represents the true value of each quantity being measured (later we adapt this model so that $x_p$ represents $h$ and $k$).
$\boldsymbol{\phi}$ is the set of hyperparameters that determine the features or shape of the prescribed analysis model.   
To infer these population hyperparameters, we must specify the priors for the hyperparameters or the hyperpriors, $p(\boldsymbol{\phi})$.
Once this multi-level model is applied to a sample of measurements, both the population's parameters and the true parameters for each of the population members can be inferred simultaneously.   
The measured properties ($d_p$) are related to the true properties ($x_p$) and the measurement uncertainties ($\sigma_p$) by $p(d_p|x_p,\sigma_p)$.
As a result, the HB model allows us to characterize the true parameter values and population level parameters while using the information contained in the measurements and their uncertainties.

The general form for the posterior for the hyperparameter vector ($\boldsymbol{\phi}$), where $D$ represents the number of measurements that make up the dataset ($d_{p}$), is given by:
%
\begin{equation}\label{eq:gen_form_post}
 p(\boldsymbol{\phi} | \mathrm{\mathbf{x_{p}},\mathbf{\sigma_p}} ) \propto p(\boldsymbol{\phi}) \prod_{p=1}^{D} \int dx_p \, p(x_p | \boldsymbol{\phi} ) p(d_p | x_p, \sigma_p)
\end{equation}

Next, let us consider a simplified HB model where each measurement $d_p$ is drawn from a normal distribution centered on the true value $x_p$ with measurement uncertainty $\sigma_p$
\begin{equation}\label{eq:prob_of_data_given_measurement}
 p(d_p|x_p,\sigma_p) \sim \,\mathrm{Normal_{\,d_p}}(x_p,\sigma_p^2).
\end{equation} 
Here, the `` $\sim$ " can be read as ``is distributed as", common notation for statisticians.
At the ``mid-level" of the hierarchical model, we assume that the population of true values, $x_p$'s, can be parameterized by a Gaussian mixture model, where each component of the population model has mean zero and $N_{m}$ is the number of mixture components. 
\begin{equation}\label{eq:anal_model}
 p(x_p | \boldsymbol{\phi}) = \, \sum_i^{N_m} f_i \,\mathrm{Normal_{\,x_p}}(0,\sigma_i^2), 
\end{equation}
Each component contributes a fraction $f_i$ of the population, so
\begin{equation}\label{eq:fraction_sum_1}
\sum_i^{N_m} f_i = 1.  
\end{equation} 

$\boldsymbol{\phi}$ then represents all of the $f_i$ and $\sigma_i$ values\footnote{Note that $\sigma_i$ is a hyperparameter that partly describes the underlying population distribution along with $f_i$, where $\sigma_p$ is the measurement uncertainties of the observable quantity.}.
If we assume a common Gaussian mixture model prior for each $x_p$ as shown in Equation \eqref{eq:anal_model} and Gaussian measurement error as shown in Equation  \eqref{eq:prob_of_data_given_measurement}, then our hierarchical model can be mathematically described as Equation \eqref{eq:gen_form_post} adapted to our specific analysis:
\begin{equation}\label{eq:full_HB_model}
\begin{split}
 & p(f_i, \sigma_i  | d_p, \sigma_p) \propto p(f_i, \sigma_i ) \\
 & \times \prod_{p=1}^{D} \int dx_p \, \sum_{i=1}^{N_m} f_i \,\mathrm{Normal_{\,x_p}}(0,\sigma_i^2) \,\mathrm{Normal_{\,d_p}}(x_p, \sigma_p^2).
 \end{split}
\end{equation}

Moving the integral inside the summation and exploiting the symmetry of the Gaussian distribution, we get
\begin{equation}\label{eq:full_HB_derivation}
\begin{split}
 & p(f_i, \sigma_i  | d_p, \sigma_p) \propto p(f_i, \sigma_i ) \\
 & \times \prod_{p=1}^{D} \left[ \sum_{i=1}^{N_m} f_i \int dx_p \, \mathrm{Normal_{\,x_p}}(0,\sigma_i^2) \,\mathrm{Normal_{\,x_p}}(d_p, \sigma_p^2) \right]. 
 \end{split}
\end{equation}
In Equation \eqref{eq:analytic_full_HB}, we extend the limits of the integral to infinity in order to develop an analytic approximation to our hierarchical model that is accurate when $\sigma_i \textless 1$, $\forall;\in [1,N_{m}]$ (i.e. allowing the underlying model to assign eccentricities \textgreater \,1).
\begin{equation}\label{eq:analytic_full_HB}
\begin{split}
 & p(f_i, \sigma_i | d_p, \sigma_p) \propto  p(f_i, \sigma_i ) \\
 & \times \prod_{p=1}^{D} \left[ \sum_{i=1}^{N_m} \frac{f_i \exp\left[ -d_p^2/(\sigma_p^2+\sigma_i^2) \right]}{\sqrt{2\pi (\sigma_p^2+\sigma_i^2)}} \right]. 
 \end{split}
\end{equation}
We discuss how we modify this derivation when evaluating our model numerically, applied to the eccentricity distribution in \textsection \ref{sec:method:priors}.   
The posterior distribution for the hyperparameter vector is conditional on all observations. 
The posterior modes and credible intervals can be calculated from Equation \eqref{eq:full_HB_derivation} using MCMC or estimated analytically based on Equation \eqref{eq:analytic_full_HB}. 
Recent applications of HB modeling applied to other \textit{Kepler} observations include \ct{Morton14}, \ct{Rogers14}, \ct{Wolfgang14}, and \ct{DFM14}.  

\subsection{Applying the Hierarchical Model to Eccentricity Measurements}
\label{sec:method:hkdata}

Next, we tailor the above model to the eccentricity distribution.
The set of projected eccentricity measurements $h$ and $k$ for each planet candidate become the $x_p$'s described in \textsection 3.1. 
We assume that each true value of $h$ and $k$ is drawn from a distribution that is a mixture of $N_m$ normal distributions (the analysis model), where each mixture component contributes a fraction $f_i$, is centered on zero, and has a standard deviation $\sigma_i$.
Thus, the hyperparameters $\phi = \left\{ f_1,..., f_{N_{m-1}}, \sigma_1, ..., \sigma_{N_m} \right\}$ describe the underlying population's distribution of $h$ and $k$'s.  
Since fractions sum to one, $f_{N_{m}}=1-\sum_{i=1}^{N_{m-1}}f_{i}$.

The values of $h$ and $k$ provide an alternate parameterization for the eccentricity ($e$), and the argument of periastron ($\omega$).
We assume that the orientations of planetary systems' pericenter directions ($\omega$) will be randomly distributed with respect to the direction towards Earth, i.e., $\omega$ is uniform random [$0,2\pi$].
Thus, the prior probability distribution for each planet's $h$ and $k$  has radial symmetry.
While this is an excellent general model for planets, it is an approximation for our sample of planet candidates since 1) the geometric transit probability and occultation probability depend somewhat on $\omega$ for eccentric orbits and 2) the detection probability of both the transit and occultation depends on the transit and occultation durations and thus the eccentricity and pericenter direction, and the occultation duration also depends on the orbital period.  
We will discus these issues further in \textsection \ref{sec:Disc}.
Results of this analysis can be found in \textsection 4 and \textsection \ref{sec:results:occs}.  

\subsection{Evaluating the Hierarchical Model}

We sample from the posterior using MCMC.
To calculate Markov chains we use the publicly-available code Just Another Gibbs Sampler \citep[JAGS;][]{Plummer03}.
JAGS uses Gibbs sampling when possible, and otherwise reverts to standard random walk Metropolis\textendash Hastings.
We simultaneously sample from both the posterior distributions for the population parameters and the posterior predictive distributions for each observable.  
We compare the within-chain variance to the between-chain variance and evaluate the Gelman-Rubin ($\hat{R}$) ratio to test for non\textendash convergence, and accept chains with an $\hat{R} < 1.01$. 
We also look at the autocorrelation function for the Markov chains and accept cases that have a zero crossing at a lag of $\leq 5$. 
The exact JAGS input model used in our study can be found online\footnote{http://www.astro.ufl.edu/\textrm{$\sim$mshabram/jags\_model/eccmodel.txt}}.

\section{Results}

\subsection{Prior specification}
\label{sec:method:priors}

We consider three different analysis models and calculate posteriors for each using simulated data to test the accuracy and robustness of our method. 
The three analysis models used for $x_p$'s in our calculation are (i) a single Gaussian ($N_{m}=1$), (ii) a two-component Gaussian mixture ($N_{m}=2$), and (iii) a three-component Gaussian mixture ($N_{m}=3$).
  
We use these same three models both to analyze the data and to generate simulated observations. 
In each model, the population parameters, also known as hyperparameters, ($\phi$) are a union of the set of dispersions for each mixture component ($\sigma_{i}$'s) and the set of fractions of planets associated with each of the mixture components ($f_{i}$'s).
In each case, each mixture component is a Gaussian centered at zero and represents a unique population. 
 
When evaluating the eccentricity parameter space, we take our priors for $h$ and $k$ to be a mixture of Gaussian distributions, each with zero mean but truncated such that $e = \sqrt{h^2 + k^2} \textless 1$.  
\changes{Following truncation, the prior for $h$ and $k$ is renormalized, so that the total probability integrates to unity.}
The truncation accounts for the selection effect of not detecting planets on hyperbolic orbits ($e \textgreater 1$) as any such planets are not bound to their host systems and do not transit more than once.
A Rayleigh distribution can also be parameterized as the square root of the sum of squared Normal distributions with zero mean, where the variance of each component is equivalent to the Rayleigh parameter.  
Thus, the prior population distribution for $e$ is a truncated Rayleigh distribution for $N_{m}=1$ and can be visualized as a mixture of truncated Rayleigh distributions of $N_{m}\geq2$. 
Choosing a Rayleigh distribution for the eccentricity distribution is physically motivated by the fact that it naturally arises for exoplanets on circular orbits and subjected to a series of many normally distributed small random perturbations to its orbit. 
It is therefore a common distribution ``shape" used for eccentricity distributions (e.g., \citet{Moorhead11}, \citet{Fabrycky14}).
We justify its superiority over a Beta distribution \cp{Kipping14b} in \textsection 4.2.2. 

Calculating a posterior probability distribution function (PDF) from a HB model also requires specifying a prior probability distribution for the population parameters ($\mathbf{\phi}$). 
This is known as the \textit{hyperprior}.  
Our hyperparameters are the dispersions $\sigma_{i}$'s ( for each mixture component) and the associated mixture fractions.  
We assume a uniform prior for the dispersions of each mixture component between 0 and 1.  
The mixture component fractions follow a Dirichlet distribution with the concentration parameter set to 1 (e.g., no component is given special weight).
This is the multidimensional generalization of the Beta distribution. 
The Dirichlet distribution forces the sum of the mixture component fractions to equal one.  

\subsection{Validating the Hierarchical Model}
\label{sec:results:ValidatingHBM}

Since the true distribution parameters for the synthetically generated datasets are known, analyzing these simulated observations with our hierarchical model allows us to directly compare the output population parameters and the input population parameters.
We are also able to test the sensitivity of the posterior to the chosen analysis model.

We expect to see variations in the ability of a given analysis model to recover the input model's parameters. 
For instance, if the analysis model is the same as the model used to generate the simulated observations, then we expect to be able to recover the input population parameters, within the limits of measurement uncertainties and Monte Carlo error. 
However, if the analysis model is different than the model used to generate the data, there could be larger differences between the posterior predictive distributions for the eccentricity distribution and the actual distributions used to generate the data. 
If we can identify an analysis model that is relatively insensitive to the model that was used to generate the simulated observations, then we can increase our confidence in the robustness of the procedure when applying our hierarchical model to a real dataset. 

\subsubsection{Generation and Analysis of Simulated Data}

We are also interested in understanding the effect that the quantity and quality of the data has on our inference. 
It is important that we choose an analysis model that is relatively robust to model misspecification, so we can be confident when applying the HB model to real data.
To accomplish this, we generate several simulated datasets varying the number of planets in the sample, the simulated measurement uncertainties, or both.
For each pair of generative model and analysis model, we analyze four datasets of different qualities.
We summarize each in Table \ref{table2:models}. 
Datasets labeled ``good" (``half") consist of $50$ ($25$) planets with measurement uncertainties of $0.04$ and $0.08$ for $h$ and $k$ respectively.\footnote{The uncertainty in the phase offset of the transit is typically smaller than that of the occultation and transit duration ratio, thus the eclipse data constrain $h$ with more precision than $k$.} 
These datasets are designed to be similar to our actual transit and occultation dataset for both $h$ and $k$.  
Datasets labeled ``better" (``best") contain $50$ ($500$) planets with measurement uncertainties of $0.001$.  
The mixture fractions and dispersions used to generate the synthetic datasets are the following: for a single Gaussian distribution, labeled as ``R1" in Table \ref{table3:KSstats}, $f=1.0$ and $\sigma=0.3$, for a two-component Gaussian mixture model (``R2"), $f_{1}=0.7$, $f_{2}=0.3$, $\sigma_{1}=0.05$, and $\sigma_{2}=0.3$, and for a three-component Gaussian mixture model (``R3"), $f_{1}=0.6$, $f_{2}=0.3$, $f_{3}=0.1$, $\sigma_{1}=0.05$, $\sigma_{2}=0.2$, and $\sigma_{3}=0.5$.
We generate 20 datasets for each pair of data quality and generative model in order to quantify Monte Carlo error.  
The goal of this particular experiment is to identify an analysis model for the $x_p$'s that performs well for a variety of plausible distributions used to generate simulated data. 

\subsubsection{Results for Synthetic Data}

First, we validate our HB model using the same model for the analysis as used to generate a simulated dataset.    
Next, we consider the results of applying an analysis model that differs from the model used to generate the data. 
\changes{The purpose of making these comparisons is to identify an appropriate analysis model, balancing the need for flexibility with the desire to minimize model parameters.  By analyzing a variety of simulated datasets, we develop intuition for how different models perform, prior to analyzing the actual data.  Since the true distribution of exoplanet eccentricities likely differs from any of our analysis models, it is important to analyze data sets generated under alternative models, so as to test the robustness of our approach.

When we use the same analysis model and generative model, it would be possible to compare the true model parameters to the posterior distribution for the model parameters.  However, most of our comparisons involve different analysis and generative models.  In these cases, it is not possible to compare the true model parameters to the posterior distribution for the model parameters.   Instead, we compare the predictive posterior distribution for the eccentricity distribution (i.e., the distribution of interest).  We use the K-S distance to measure how the predictive posterior distribution for eccentricities  under each analysis model compares to the true eccentricity distribution used to generate the simulated data.}  
Table 3 shows the median Kolmogorov-Smirnov (K-S) distance between each simulated dataset's true $h$ and $k$ distribution and the posterior predictive distribution for $h$ and $k$ based on $20$ simulations of a particular hierarchical model from Table 2 (see \textsection 4.2.1 for a list of the chosen ``true" eccentricity distribution values used in our study).

\begin{figure*}
\epsscale{0.7}
\plotone{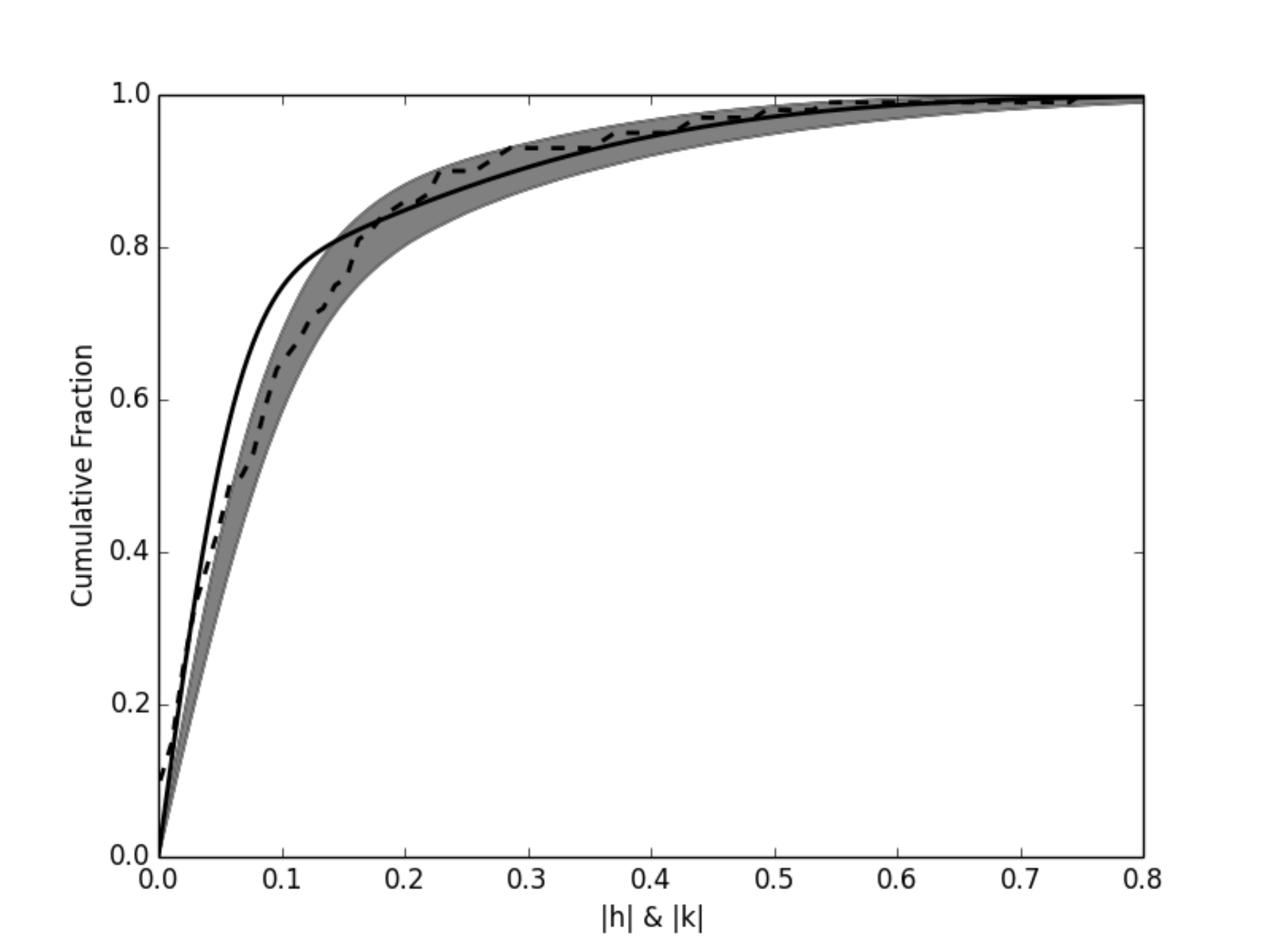}
\caption{Cumulative distributions of $|h|$ and $|k|$.  The solid black curve is the true distribution from which the simulated planet's $h$ and $k$ values are drawn.  The dashed black curve is the cumulative distribution for one simulated dataset (``R2", ``good"; $f_{1}=0.7$, $f_{2}=0.3$, $\sigma_{1}=0.05$, and $\sigma_{2}=0.3$, see \textsection4.2.1) that includes simulated observational uncertainties.   \changes{The gray shaded region is the 68.3\% credible interval for the CDFs of the posterior distribution for the population parameters of the intrinsic distribution of $|h|$ and $|k|$ (i.e., without intrinsic uncertainties).  This is calculated once the simulated observations have been analyzed using the same two-component Gaussian mixture model as was used to generate the data.}  The two-component Gaussian HB model does a good job of capturing the true distribution for datasets generated with a two-component Gaussian mixture.  
}
\label{fig:cumulative}
\end{figure*}

We illustrate an example case in Figure \ref{fig:cumulative}, by showing cumulative distributions of $|h|$ and $|k|$. 
The solid black curve is the true distribution from which the simulated planet's h and k values are drawn.  
The dashed black curve is the cumulative distribution for one simulated dataset (``R2", ``good"; $f_{1}=0.7$, $f_{2}=0.3$, $\sigma_{1}=0.05$, and $\sigma_{2}=0.3$, see  \textsection 4.2.1) that includes simulated observational uncertainties.  
\changes{The gray shaded region is the 68.3\% credible interval for the CDFs of the posterior distribution for the population parameters of the intrinsic distribution of $|h|$ and $|k|$ (i.e., without intrinsic uncertainties).  This is calculated once the simulated observations have been analyzed using the same two-component Gaussian mixture model as was used to generate the data.}

The posterior predictive distributions are generated from the posteriors for the hyperparameters obtained from applying the hierarchical model to each simulated dataset.  
Each column of Table 3 represents comparison results for posteriors calculated using an analysis model with one-, two-, or three-components in the Gaussian mixture model respectively, for each eccentricity distribution.
The analysis model names are described in Table 2. 
Each row of Table 3 gives results for a specific generative model and data quality.  
R1, R2, and R3, indicate one, two and three component models for generating the simulated observations, respectively.  

If the model is working properly, we expect to get posterior distributions for the population's parameters that are consistent with values used to generate the data.  
Indeed, we find K-S distances are between $\sim$ 0.05 to 0.1 for these cases. 
Since models have different parameters (even when they are represented by the same variable names), the most appropriate way to compare the performance is based on the posterior predictive distribution for the population of measurements.  In this case, the K-S distances between the posterior predictive distributions for the HB model and actual model are $\sim$ 0.1 to 0.2.  

For simulated datasets with smaller measurement uncertainties, we find that the K-S distance between the posterior predictive distribution and associated simulated data for $h$ and $k$ is similar for analysis models that have at least the same number for mixture components or more.  
Additionally, for several combinations of analysis and generated models, we note that the $N_{m}=2$ analysis model results in a smaller K-S distance to the R3 data than the $N_{m}=3$ analysis model.
This is likely due to the greater flexibility of the $N_{m}=3$ model and finite number of measurements, i.e., the three-component model ``over-fits" the discrete dataset.   
We found that the $N_{m}=2$ analysis model did a better job overall at recovering the predictive distribution for the simulated datasets across all versions for simulated data.  

\begin{figure*}
\epsscale{0.7}
\plotone{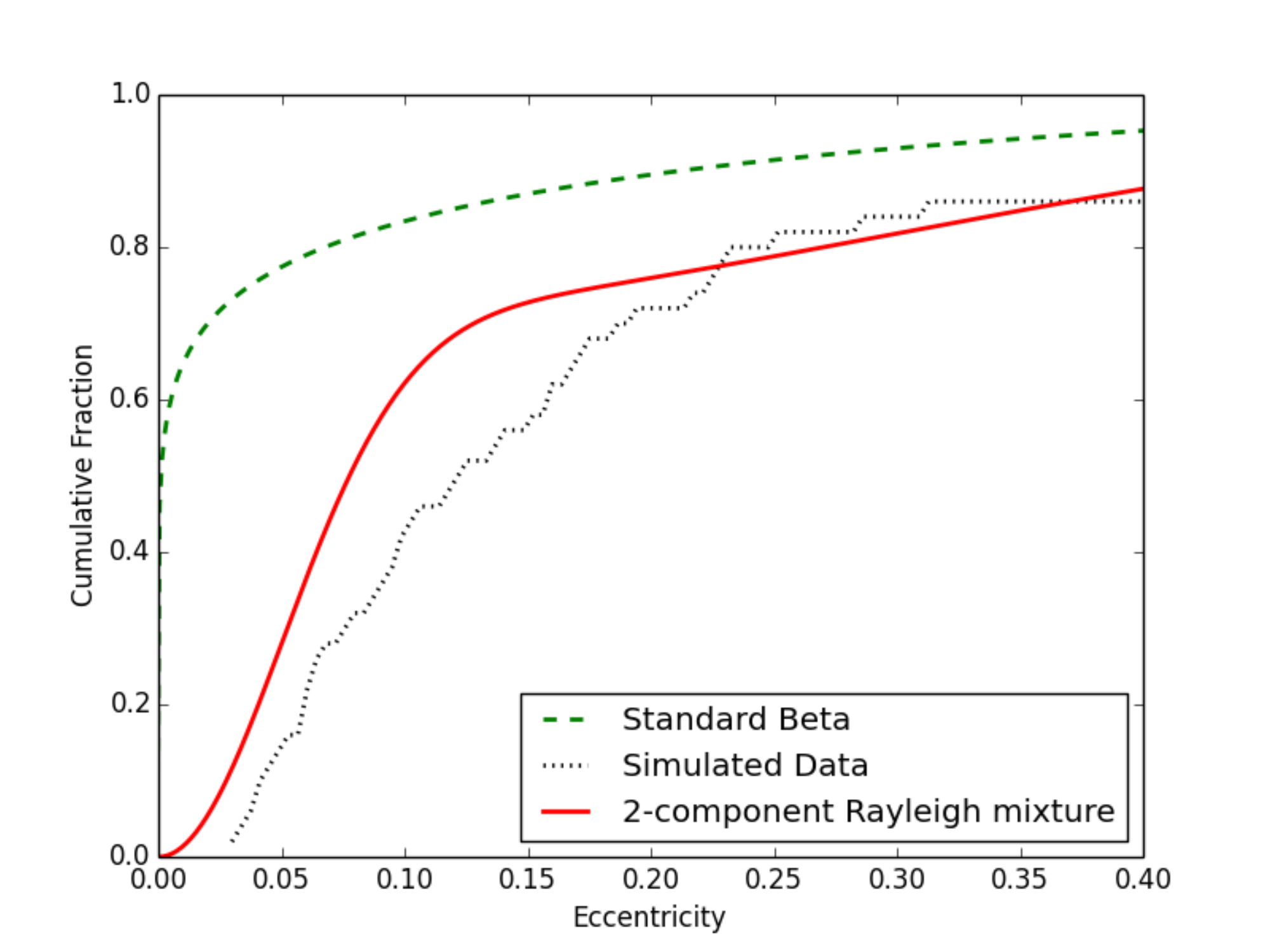}
\caption{ Results of an HB model that parameterizes the eccentricity as a standard Beta distribution. We investigate using a Beta distribution analysis model on an "R2", "good" (see Table 2) simulated dataset.  
\changes{A cumulative distribution of} the simulated eccentricity data are shown as the dotted black curve.  The distribution generated using a two-component Gaussian mixture model for $h$ and $k$ values (e.g., $f_{1}=0.7$, $f_{2}=0.3$, $\sigma_{1}=0.05$, and $\sigma_{2}=0.3$, as described in \textsection 4.2.1) is shown in red. 
The dashed green curve is a \changes{cumulative} Beta distribution, Beta($\alpha$, $\beta$), plotted using the posterior modes for $\alpha$ and $\beta$, ($\alpha = 0.11\pm^{0.04}_{0.02}$, $\beta = 1.73\pm^{0.85}_{0.24}$) for this HB model.  
The Beta distribution erroneously predicts a strong peak near $e=0$ and under predicts the frequency of larger eccentricities.     
}
\label{fig:Nm2_R2_Beta}
\end{figure*}

Some authors advocate parameterizing the eccentricity distribution as a Beta distribution, $e\sim$ Beta($\alpha$, $\beta$) \citep[e.g.,][]{Kipping14b}.  
Therefore, we also investigate using a Beta distribution analysis model using one (``R2", ``good") simulated dataset (see Table 2).  
In this model set up, $\alpha$ and $\beta$ become the hyperparameters (population level parameters) that we wish to infer.  
We use a Gamma distribution with $k=2$ and $\theta=1$ as the prior probability distribution for $\alpha$ and $\beta$. 
Figure \ref{fig:Nm2_R2_Beta} shows the results of this HB model as eccentricity vs. \changes{cumulative fraction}.  
The simulated eccentricity data are shown as the dotted black curve.  
The true eccentricity distribution generated using a two-component Gaussian mixture model for $h$ and $k$ (e.g. $f_{1}=0.7$, $f_{2}=0.3$, $\sigma_{1}=0.05$, and $\sigma_{2}=0.3$, as described in \textsection 4.2.1) is shown in red. 
The dashed green curve is plotted using the posterior modes for $\alpha$ and $\beta$, ($\alpha = 0.11\pm^{0.04}_{0.02}$, $\beta = 1.73\pm^{0.85}_{0.24}$) for this HB model.  
The K-S distance between the ``R2," ``good" distribution (red) used and the distribution using the posteriors modes of $\alpha$ and $\beta$ (dashed-green) is $0.5$, which is in support of these being two distinct distributions.  
Our results indicate that the standard Beta distribution is a poor choice for an analysis model to parameterize the eccentricity distribution as it erroneously predicts a strong peak near $e=0$ and under predicts the frequency of larger eccentricities.

\changes{In principle, we could model the transit and occultation times and durations directly, instead of $h$ and $k$.  However, this would increase the computational complexity.  Even taking advantage of our approximations, the calculations presented represent a significant computational investment ($\sim$2.4 CPU months).  Modeling $h$ and $k$ also facilitates deriving analytical expressions for testing the algorithms.  By modeling the projected eccentricity, $h$ and $k$, we were able to thoroughly test both the code and algorithm on real and simulated data sets.}

\section{Results for \textit{Kepler} planet candidates with occultations}
\label{sec:results:occs}

\begin{figure*}
\epsscale{0.7}
\plotone{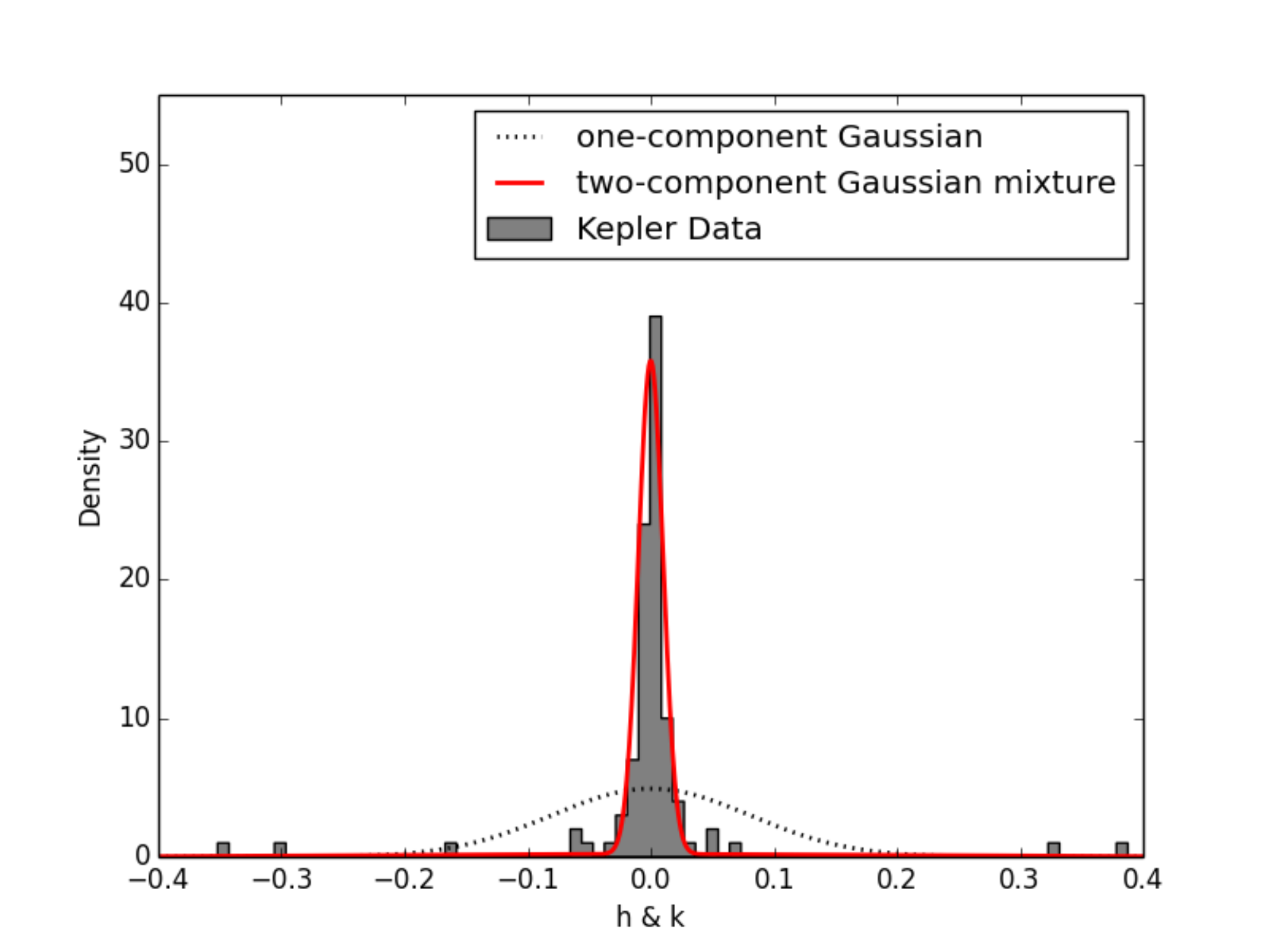}
\caption{A histogram of the $h$ and $k$ dataset are shown in grey. The dotted black curve is a one-component Gaussian distribution using the posterior mode for the dispersion obtained from an HB model that uses a one-component Gaussian mixture model.  Shown in red is a two-component Gaussian mixture model using posterior modes for the mixture fractions and dispersions obtained from an HB model that uses a two-component Gaussian mixture model.  The black model does a poor job at capturing the shape of the distribution.  The red model captures the peaked nature of the true distribution while also allowing for a smaller number of measurements far from the central peak.}
\label{fig:Nm2_all_hk}
\end{figure*}

We calculated posteriors of one-, two-, and three-component Gaussian mixture models applied to real \textit{Kepler} transit and occultation data (see \textsection 2 for description of dataset).
Figure \ref{fig:Nm2_all_hk} shows a histogram of the observed $h$ and $k$ values from Table 1 (shown in grey). 
Since we are assuming the argument of periastron ($\omega$) is random, $h$ and $k$ are equivalent, or drawn from the same distribution. 
A Gaussian distribution using population parameters from the posterior mode for the dispersion for a one-component model is overplotted (shown as the dotted black curve).  
This one-component model does a poor job at capturing the shape of the distribution because it struggles to match the moderate eccentricity outliers.  
A two-component Gaussian mixture model using population parameters from the mode of the 2D marginal posterior for the mixture fraction and dispersions is shown in red.  
This model captures the peaked nature of the observed distribution as well as the small number of measurements away from the peak.  
This suggests that two populations can explain the eccentricity distribution of our sample, although in \textsection 4.4.2 we show with synthetic data that using a two-component Gaussian mixture model is optimal for the present dataset.  
We also consider using a three component Gaussian mixture model and find that only two of the three components can be constrained given the quantity and quality of the \textit{Kepler} transit and occultation dataset, suggesting that the available data are not able to indicate the presence of a third population, or that a third population may not exist.  

\begin{figure*}
\epsscale{0.7}
\plotone{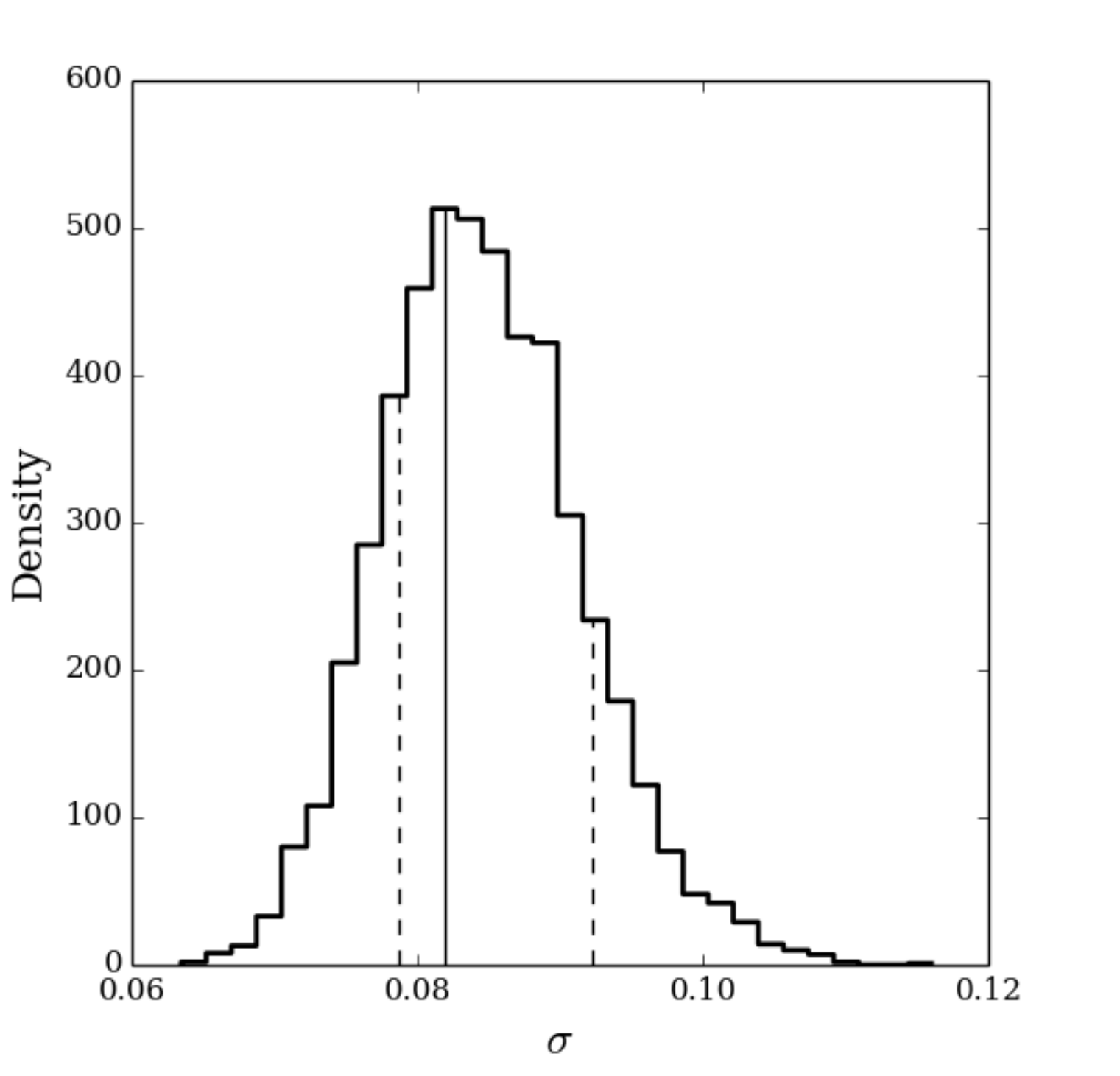}
\caption{Posterior distribution for the dispersion of a Gaussian model for $h$ and $k$ applied to our full \textit{Kepler} short-period planet candidate transit and occultation dataset.  The $68.3\%$ credible intervals about the mode are shown as dotted black lines, and the mode is shown as a vertical solid black line.  A one-component Gaussian mixture model is insufficient at characterizing the eccentricity distribution of our sample of planet candidates from \textit{Kepler}.  
}
\label{fig:Nm1_all}
\end{figure*}

The posterior distribution for the dispersion of true values of $h$ and $k$ assuming a one-component Gaussian model is displayed in Figure \ref{fig:Nm1_all}, which is based on our full dataset for planet candidates with both \textit{Kepler} transit and occultation measurements.  The posterior mode for the dispersion is $\sigma=0.081 \pm^{0.014}_{0.003}$. We use this value as the dispersion for our one-component Gaussian population model shown as the dotted black curve in Figure \ref{fig:Nm2_all_hk}. 

Next, we investigate joint posterior distributions for a two-component Gaussian mixture model applied to our full \textit{Kepler} transit and occultation dataset.  
These results are shown in Figure \ref{fig:Nm2_all_corner}, where the panels on the diagonal show the marginalized posterior distribution for the population parameters: $\sigma_{low}$ the lesser value of $\sigma_{1}$ and $\sigma_{2}$,  $\sigma_{high}$ the greater value of $\sigma_{1}$ and $\sigma_{2}$,  and $f_{low}$, the weights for the mixture component ($f_{high} = 1 - f_{low}$).
The use of $\sigma_{low}$ and $\sigma_{high}$ instead of $\sigma_{1}$ and $\sigma_{2}$ (and corresponding fractions) is helpful for visualizing the results, since our model has symmetry under exchanging ($\sigma_1,f_1$) and ($\sigma_2,f_2$).    
The off-diagonal panels show posterior samples and contours for the 68.3\% credible interval of the two-dimensional marginal posteriors for each parameter pair.     
The fact that ($\sigma_{low}$, $f_{low}$) and ($\sigma_{high}$, $f_{high}$) form two distinct clusters demonstrates the value of a two-component (two population) model for the eccentricity distribution of our sample of \textit{Kepler} planet candidates.  

As expected, the uncertainties in measurements of $k = e \sin \omega$ are much greater than the uncertainties in the measurements of $h=e \cos \omega$. 
We note that in our sample, the $k$ values are more tightly clustered around zero than $h$.  
Therefore, we investigated if excluding these values significantly impacts our results. 
By doing this we are decreasing our effective sample size, but maintaining the number of measurements with small uncertainties.  
When running our simulations without $k$ values, we get the following results: using a single Gaussian model $\sigma=0.074 \pm^{0.016}_{0.003}$, and when using a two-component Gaussian mixture model, the marginal posterior modes for the mixture fractions and dispersions are $f_{low}=0.93 \pm ^{0.029}_{0.051}$, $\sigma_{low}=0.003 \pm^{0.010}_{0.001}$, and $f_{high}=0.07 \pm^{0.062}_{0.019}$, $\sigma_{high}=0.187 \pm^{0.547}_{0.028}$, respectively.  
These values differ  from the values obtained using the full $h$ and $k$ planet candidate dataset by $\sim$ $9.0\%$, $4.4\%$, $107.7\%$, $44.4\%$, and $16.2\%$ for $\sigma$, $f_{low}$, $\sigma_{low}$, $f_{high}$, and $\sigma_{high}$ respectively. 
Each of these overlaps the $68.3\%$ credible interval for the same parameters with the full dataset.  
The most notable difference is for $\sigma_{low}$, which we estimate to be $0.03$ when using $k$ alone, but $0.01$ when including both $h$ and $k$ observations. 

\subsection{Correlation of The Eccentricity Distribution with Star and Planet Properties}

Another goal of this study is to investigate potential correlations between the eccentricity distribution and planet or host star properties. 
Specifically, we consider whether the planet candidate eccentricity distribution is correlated with stellar metallicity, host star effective temperature, planet radius, or orbital period. 
Values for each planet candidate are obtained from the \textit{Kepler} Star Properties Catalog as reported at Exoplanet Archive updated December 2013 and revised February 2014 \cp{Buchhave12, Huber14}. 
The majority of the planets have effective temperatures and stellar metallicity values obtained via KIC photometry, however 10 planets have spectroscopically derived values.
Given the relatively small sample size, we focus on comparing the distribution of planet candidates with large and small values for each parameter. 
We sort the planet candidates in our dataset from largest to smallest values of a given property, and then create two sub-samples of the original population. 
Unless otherwise specified, we divide the data in half to maximize the statistical power when comparing the two samples and to avoid introducing an additional parameter specifying the dividing point between the high and low subsets.
We analyze each subset as described in section \textsection 3.

Initially, we evaluate each subset of data using an HB model with a one-component Gaussian distribution for $h$ and $k$, as we did before for the full dataset.  This is shown in Figure \ref{fig:Nm1_halves_hk}, where we have applied the HB model to the small-valued (blue) and large-valued (red) halves of the \textit{Kepler} occultation data, sorted by (a) stellar effective temperature,  (b) planet radius, (c) orbital period, and (d) stellar metallicity.  
The histograms of the posterior distribution for the dispersions for stellar effective temperature (a) and orbital period (c) suggest that the two subsets do not come from significantly different distributions if we assume the eccentricity distribution is described by a simple Rayleigh distribution.  
However, for planet radius (b), and stellar metallicity (d), the differences in the posteriors for the dispersion suggests that the two subsets may have different eccentricity distributions.  
This provides motivation to consider more complex models for correlations between the eccentricity distribution and stellar and planet properties.

Next, we look at posterior distributions based on applying the HB model using a two-component Gaussian mixture for the analysis model applied to the small-valued half or large-valued half of the data subsets, again based on sorting by (a) stellar effective temperature,  (b) planet radius, (c) orbital period, and (d) stellar metallicity. 
Figure 7 shows posterior distributions for the small-valued half of data (blue and green clusters), and large-valued half (red and orange clusters).  
The two groups of clusters represent samples of the posterior distribution for the hyperparameter vector, in this case for $\sigma_{low}$ and $f_{low}$ (top left group of clusters in each sub-plot), and $\sigma_{high}$ and $f_{high}$ (bottom right group of clusters in each sub-plot). 
The data are plotted with the vertical axis representing the low value of the mixture fraction, $f_{low}$, in green and orange, and, $f_{high}$, in blue and red for the two subsets of sorted data shown (small and large).  
The contours correspond to 68.3\% credible intervals.  
In this plot we can compare the two high and low value subsets to the full sample to check for correlations with each parameter.  

Interestingly, the posteriors of the mixture fractions for the planet candidates with larger planet radii are consistent with 0 and 1 for planet radius (b) and host star metallicity (d), indicating only one population is required to accurately model the eccentricity distribution for this subset of planet candidates.
When modeling the eccentricity distribution of a sample of planet candidates with host star metallicities less than 0.023 dex (median [Fe/H] of our full sample), we find a one-component mixture model is sufficient. 
On the other hand, planet candidates of host star metallicities above 0.023 dex are better modeled with a two-component mixture model for $h$ and $k$. 
We find a similar, but weaker correlation between eccentricity distribution and planet radius.  Planet candidates with radii smaller than 10.6 $R_\oplus$ are better modeled with a two-component mixture model, while a one component mixture model is favored for planet candidates with radii above 10.6 $R_\oplus$.  
There is not a strong correlation between planet radius and metallicity in our sample.  
Further, we verified that the subsamples based on metallicity and planet radius are distinct from each other.
 
Next, we consider whether the data can constrain more complex models that allow for a flexible choice of the break point between the two subsets of planet candidates, rather than fixing the break point to divide the dataset in half.  
We choose to investigate a more flexible model for orbital period first, because the measurement uncertainties for orbital period are negligible.
We allow the period break point to be a free parameter in a single HB model where each subset of planet candidates are modeled with a two-component mixture model as before.  
Instead of fixing the break point near the median and dividing the data into equal sized subsets, we place a uniform prior on the period break point.  
Figure \ref{fig:period_break} shows the marginal posterior distribution for the period break point from analyzing the actual dataset.
The marginal posterior for the period break has peak values that are clustered near the minimum and maximum of the period values in our sample.  
This indicates that the model favors period breaks causing one data subset to have so few observations that the population parameters from one subset are minimally constrained. 
We conclude that the present dataset is not able to usefully constrain this more complex model. 
Therefore, we do not attempt to apply a similar model allowing for two eccentricity distributions with an unknown break in terms of the host star metallicity or planet radius, since their measurement uncertainties are much larger.  

\section{Summary of Results}
 
We investigated the eccentricity distribution for a sample of short-period single-planet candidate systems from \textit{Kepler} that are detected in both transit and occultation.  
We demonstrated that HB models are well-suited for characterizing the eccentricity distribution using transit and occultation data.
We modeled the distribution of $h$ and $k$ as coming from either a single Gaussian distribution with zero mean or a mixture of Normal distributions.  After testing our hierarchical model on a suite of simulated datasets and analysis models with one, two, or three mixture components, we find that a two-component mixture model ($N_{m}=2$) performed well in all cases considered, including simulated datasets generated using a three-component mixture model.  Thus, the two component mixture model is a robust analysis model for our hierarchical model applied to \textit{Kepler} transit and occultation data.  Additionally, we investigate the usage of a standard Beta distribution analysis model in our HB model.  
Our results indicate that the standard Beta distribution is a poor choice for an analysis model to parameterize the eccentricity distribution.

Next, we applied HB modeling to analyze a real dataset of $h=e \cos \omega$ and $k=e \sin \omega$ measurements, derived from transit and occultation measurements.  If we model the population distribution of $h$ and $k$ with a single Gaussian, then we infer a dispersion of $\sigma=0.081 \pm^{0.014}_{0.003}$.
When we applied the two-component mixture model to the full dataset,  we found $f_{low}=0.89 \pm ^{0.045}_{0.057}$, $\sigma_{low}=0.01 \pm^{0.014}_{0.002},$ and $f_{high}=0.11 \pm^{0.057}_{0.045}$, $\sigma_{high}=0.22 \pm^{0.100}_{0.026}$.
These results suggest the presence of a small population of planet candidates ($\sim 11$\%) that contain planets with a broad range of orbital eccentricities and a larger population of planet candidates ($\sim 89\%$) that contain planets on very nearly on circular orbits. 

Next, we assessed whether there is evidence for more complexity in the eccentricity distribution by considering analysis models that allow correlations between the eccentricity distribution and other planet or host star parameters.   
For the current sample of \textit{Kepler} planet candidates seen in both transit and occultation, we find interesting correlations of the eccentricity distribution with either the planet radius or the host star metallicity, but not with stellar effective temperature or orbital period.

We present evidence that host stars in our sample with higher metallicity and planet candidates with smaller radii have a more complex eccentricity distribution than stars with low metallicity and planet candidates with larger radii.  The eccentricity distribution of these more complex populations are well described by a two-component Gaussian mixture model with a zero mean, suggesting a potential physical explanation in terms of proposed planet formation models, which we will describe in more detail in \textsection \ref{sec:Disc}.  

\section{Discussion}
\label{sec:Disc}

Previous studies of the period-size distribution of \textit{Kepler} planet candidates have identified two common architectures of planetary systems:  1) Systems with Tightly-packed Inner Planets (STIPS; \ct{Lissauer11, Payne13, Boley14}) and 2) systems with a single short-period planet (often a giant planet) with either no additional planets detected or a large gap between the short-period planet and the next detectable planet \cp{Steffen12b, Steffen13, Dawson13}.  
This work focuses on a sample containing primarily isolated giant planet candidates.  
Of course, these systems may have undetected companions, particularly at larger orbital separations where the geometric transit probability is small.  

Two broad classes of mechanisms have been proposed to explain the formation of hot Jupiters.  
In both models, planets form at greater distances from their host star.  
In one model, a giant planet experiences a gradual inspiral through a gas or planetesimal disk until they halt near their present orbit (e.g., \citealt{Kley12}).  
In the other model, gravitational perturbations from another massive body (potentially another planet or a stellar companion) excite a giant planet's orbital eccentricity until its periastron distance is so small that tidal forces begin to circularize the orbit (e.g., \citealt{Rasio96, Fabrycky07b, Naoz11}).  
Several studies have assessed the relative merit of these two classes of models, making use of the observed orbital period distribution \cp{Ford06, Valsecchi14}, spin-orbit obliquity distribution \cp{Morton11, Albrecht12, Naoz12, Dawson14} and orbital architectures \cp{Steffen12a, Dawson12b, Dawson13}.  
Collectively, these studies suggest that multiple mechanisms likely contribute to the formation of hot-Jupiters.  
In this case, the two populations would likely have different distributions of orbital eccentricities, with disk migration leading to the smaller dispersion of eccentricities.  
This motivates us to consider interpreting evidence of a two-component mixture model for the eccentricity distribution in terms of two formation models.  

Our analysis of \textit{Kepler's} short-period planet candidates with occultation measurements suggests that the eccentricity distribution can be well described by a two-component mixture model, where the less abundant population of planets has a broader dispersion of eccentricities.  If the mixture components indeed translate into formation mechanisms of hot-Jupiters, then this could suggest that disk migration could be the more common formation mechanism.  
Alternatively, if tidal circularization of highly eccentric proto-hot Jupiters is sufficiently rapid, then the current eccentricities could reflect late-stage excitation of orbital eccentricities due to undetected planets.  
Of course, a complete formation theory would need to explain all observations, including the low abundance of additional planets near hot-Jupiters \cp{Steffen12b}, the final semi-major axis of hot-Jupiters (e.g., \citealt{Ford08b, Valsecchi14}), the distribution of orbital obliquities (e.g., \citealt{Fabrycky09, Morton11, Albrecht12}) and correlations between obliquity and other star and planet properties (e.g., \citealt{Winn10, Morton14}).

It is particularly interesting to compare the eccentricity distribution of our sample to that of other subsamples of the \textit{Kepler} planet candidate list, particularly subsamples dominated by smaller planets.  
Since the eccentricity affects the transit duration \cp{Barnes07c, Burke08}, the distribution of transit durations can constrain the eccentricity distribution for arbitrary sub-samples of \textit{Kepler} planet candidates \cp{Ford08a}.  
Early studies of the eccentricity distribution of \textit{Kepler's} planet candidates \cp{Moorhead11, Kane12, Plavchan12} were limited due to the uncertainty in stellar parameters.  
Transit durations combined with stellar properties from photometry \cp{Moorhead11}, high-resolution spectroscopy \cp{Buchhave12, Dawson12a} and/or flicker \cp{Kipping14a} can effectively recognize high eccentricity planets.  
However, further research is needed to obtain stellar properties precise and accurate enough to enable population studies of the more typical low eccentricity planets.  
Fortunately, one can characterize the eccentricity distribution of substantial subsets of \textit{Kepler} planet candidates, either by using stars with high quality stellar characterization (e.g., asteroseismology, \citealt{Huber13}; Ford et al.\ in prep.) or by using ratios that eliminate the dependence on stellar properties (e.g., \citealt{Kipping11}; \citealt{Fabrycky14}; Morehead et al. in prep.).  

Previous studies suggest that the typical eccentricity of planet candidates in systems with multiple transiting planet candidate systems ($\simeq~0.00-0.06$) is likely smaller than in our sample (e.g., \citealt{Fang12, Fabrycky14}).  Similarly, \ct{Wu13} and \ct{Hadden14} analyze transit timing variations (TTVs) in systems with near-resonant planet candidates.  \ct{Hadden14} report a maximum likelihood estimate of the dispersion of eccentricities of $\sigma_e = 0.018^{+0.005}_{-0.004}$.  This is significantly smaller than the $\sigma=0.08^{+0.014}_{-0.003}$ of our one-component model for the eccentricity distribution.  Their result is comparable to the $\sigma_{\mathrm{low}} = 0.01^{+0.014}_{-0.002}$ that describes nearly 90\% of planets when using our two-component model.  \ct{Hadden14} explicitly consider dividing their sample into subsets based on the estimated planet size being larger or smaller than $2.5R_{\oplus}$.  Interestingly, they find an even smaller eccentricity dispersion ($\sigma = 0.008^{+0.003}_{-0.002}$) for subset of larger planets with measurable TTVs.  This is opposite of what would be expected based on a simple comparison to either our results or planets discovered by radial velocity surveys, both of which are dominated by significantly larger and/or more massive planets.  Future observations and analyses with improved statistical methodology will be important for understanding the underlying nature of these differences.  

\subsection{Potential Biases}
We note several potential sources of bias in our characterization of the eccentricity distribution.  
First, there is a purely geometrical effect due to the fact that we analyze only planet candidates observed to both transit and occult their host stars.  
Since the geometrical transit and occultation probabilities are both functions of the orbital eccentricity and direction of pericenter, the eccentricity distribution of planets in our sample is different than the eccentricity distribution of all planets (even if we controlled for orbital separation relative to the star radius).  Mathematically, we assume a uniform distribution for the argument of periastron, $\omega$, which is true for all planets in nature, but not true for our sample.  Previous studies have suggested that the difference between the eccentricity distribution of transiting planets and all short-period planets is modest \cp{Burke08, Kipping14b}.  
For our sample, the effect will be even weaker, since we require both transit and occultation to be included in our sample and the two have opposite dependence on $e \sin \omega$.  
\changes{We confirm this by computing the distribution of $\omega$ for a simulated population generated by starting with a uniform distribution and rejecting planets that do not both transit and occult. 
We find $\omega$ to be very nearly uniformly distributed for these cases, where there is a non-zero effect but it is less that 4\%.}
Therefore, we do not account for this effect in our study.  We quantify the significance of this effect in an upcoming study by employing Approximate Bayesian Computing which can  naturally model complex selection criteria such as this (Cisewski et al. in prep).  

A second potential bias in our characterization of the eccentricity distribution is simply that the population of planet candidates we study may not be representative of all planet candidates.  
The decreasing geometric transit probability as a function of orbital period or semi-major axis inevitably leads to our sample being dominated by planet candidates with short orbital periods, similar to all population studies based on known transiting planets.  
Detection probability is also a function of the signal-to-noise of the transit and occultation.  Since the transit is typically much deeper than the occultation, the detection probability for the occultation (rather than the transit) is the dominant effect for this study.  
The occultation signal-to-noise depends on the occultation depth and duration. 
In principle, the duration depends on the density of the host star, impact parameter, eccentricity and pericenter.  
In practice, the occultation duration is most sensitive to the impact parameter.  
Thus, our sample may have excluded some planets with large impact parameters, resulting in the occultation going undetected.  
Similarly, the occultation depth depends on the effective temperature of the planet, and thus indirectly on the effective temperature of the star, the orbital distance and the stellar radius.  
Therefore, our sample is likely enriched in planets with larger radii, planets with higher effective temperatures, host stars with higher effective temperatures and smaller radii, and planets orbiting even more closely to their host star.  
If planet formation proceeds differently around more massive stars, then the eccentricity distribution for planets in our sample could deviate from the eccentricity distribution of the overall planet population.  

Furthermore, our sample could be enhanced with planets with significant eccentricity if tidal forces on planets in eccentric orbits led to substantial heating and increased thermal emission.  
Since the sample analyzed in this study consists of mostly giant planets and host stars with a single detected transiting planet, the eccentricity distribution for our sample may differ from the eccentricity distribution of smaller planets and/or planets in systems with multiple closely-spaced planets. 
These potential biases can also be viewed as opportunities to constrain planet formation and tidal theories.   
By comparing the eccentricity distribution of different planet populations, future studies can quantify how the eccentricity distribution changes with planet size, multiplicity and stellar properties.  

We anticipate several ways that future observations will allow for improvements to our analysis.  
First, we analyzed a subsample of the \textit{Kepler} planet candidates that had already been evaluated for any indication that the \textit{Kepler} Object of Interest (KOI) was actually due to an eclipsing binary star, rather than a planet (e.g., \citealt{Tenenbaum14, Bryson13}).  
Both the transit shape and comparison of the target centroid location during and out-of-transit provide powerful diagnostics for recognizing likely false positives.  
Estimates of the false positive rate are sufficiently low ($\sim10-20\%$; \citealt{Fressin13, Burke14}), that we can interpret our results in terms of the eccentricity distribution of planets.  
Nevertheless, one should be cognizant that the sample of planet candidates we analyze may include one or more false positives, such as diluted eclipsing binaries.  
In particular, our study necessarily selects planet candidates for which an occultation is measured, which may lead to an increased rate of diluted eclipsing binary false positives.  
Properly accounting for a non-zero fraction of false positives would require a significantly more complex model.  
Therefore, we leave such work for future studies.  
Alternatively, future observations of these very interesting planet candidates may identify any remaining false positives and characterize the false positive rate sufficiently well that adding further complexities to the model is not necessary.  

\subsection{Future Research}

Transit and occultation observations from future missions could lead to improved understanding of the eccentricity distribution of short-period planets as a function of host star mass and temperature.   
In particular, there could be differences between a volume-limited sample of target stars and our sample due to the target selection algorithm for the \textit{Kepler} planet search targets as well as variations in \textit{Kepler's} detection sensitivity.  
Given the \textit{Kepler} target selection criteria and detection sensitivity, most of the planet candidates we analyze are orbiting F and G stars.   
Future missions such as NASA's Transiting Exoplanet Survey Satellite (TESS) \cp{Ricker14} and ESA's PLATO \cp{Rauer14} are expected to survey a broader range of target stars and to have a simpler target selection function.  

Finally, future observations and analysis of host star properties could also result in improved characterization of correlations between the eccentricity distribution and host star properties.  In particular, a large fraction of host star metallicities used in our analysis were derived from photometric observations as opposed to higher quality spectroscopic observations \cp{Huber13}.  As metallicities derived from high-resolution spectroscopy are published for more stars in our sample, we will be able to make more robust conclusions about the potential correlation of the eccentricity distribution with metallicity.  

This paper demonstrates that it is practical to apply rigorous HB models to evaluate key dynamical properties of exoplanet populations.  
In principle, these methods can be readily generalized to provide a more accurate characterization of other aspects of the exoplanet population, such as the frequency of planets as a function of size and orbital period \cp{DFM14}, the planet radius-mass relationship \cp{Rogers14, Wolfgang14}, the distribution of mutual orbital inclinations and multiplicity, and the frequency of small planets in the habitable zone of solar-type stars \cp{DFM14}.  
\changes{A challenge for future HB analysis will be to develop rigorous model comparison techniques. 
Each unique problem is often limited by the statistical power of the data, where there is no universal technique applicable in all cases.}
In practice, the high-dimensional integration required can be computationally challenging.  
Therefore, careful thought and problem specification is needed, so as to render the necessary calculations tractable.  
Fortunately, recent collaborations between astronomers and statisticians, such as the 2013 program on Modern Statistical and Computational Methods for Analysis of \textit{Kepler} Data (SAMSI) at the Statistical and Applied Mathematical Sciences Institute have significantly enhanced the level of sophistication among exoplanet researchers.  
Forthcoming publications will describe recent efforts application of importance sampling and Approximate Bayesian computing (ABC) to enable application of HB models to more complex problems (e.g., \citealt{Rogers14, Morton14}; Cisewski et al. in prep).

\acknowledgements

\begin{figure*}
\epsscale{0.8}
\plotone{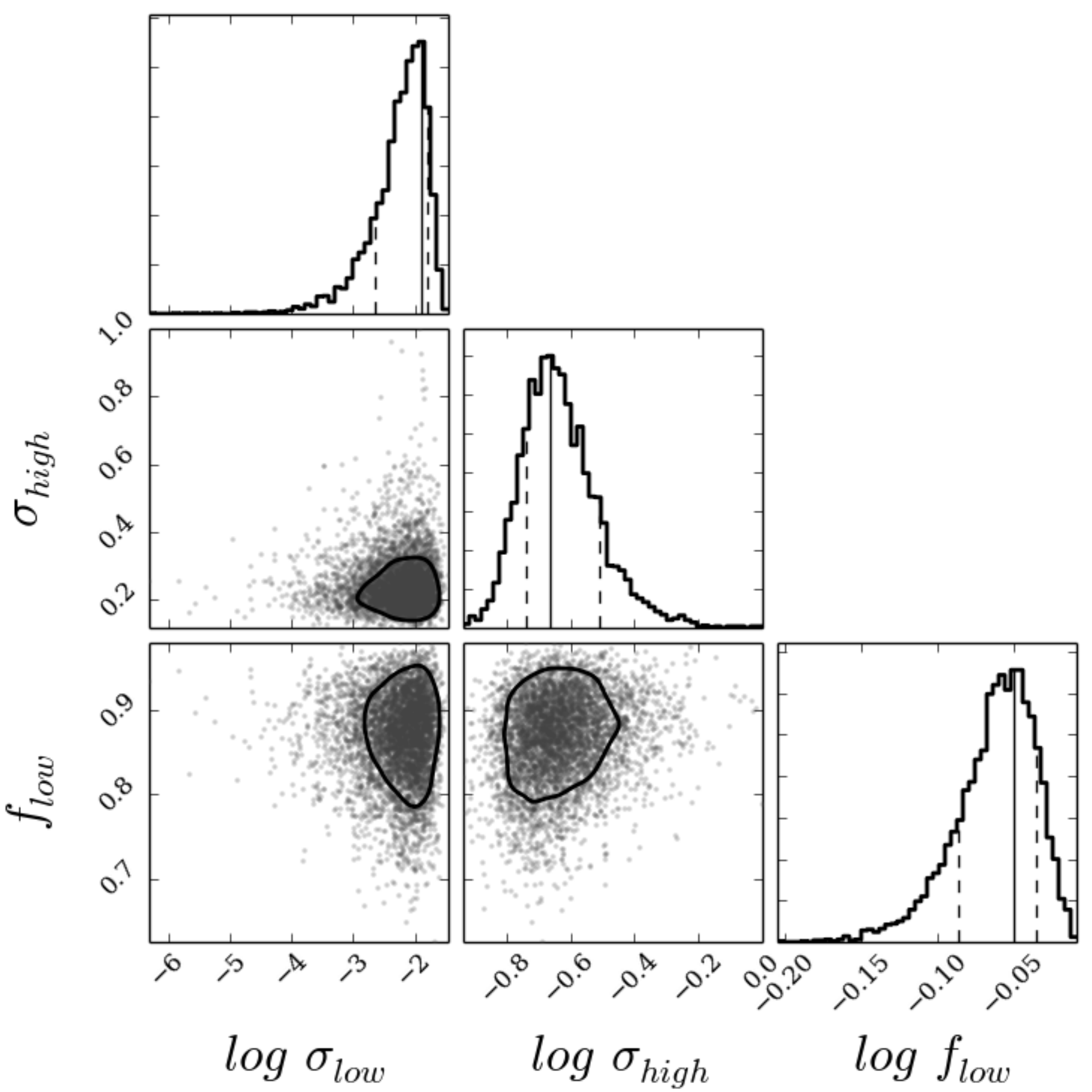}
\caption{Joint posterior distributions for a two-component Gaussian mixture model applied to our \textit{Kepler} short-period planet candidate transit and occultation dataset. 
In each panel, the data are plotted with the horizontal axis representing $\sigma_{low}$ the lesser value of $\sigma_{1}$ and $\sigma_{2}$,  $\sigma_{high}$ the greater value of $\sigma_{1}$ and $\sigma_{2}$, and $f_{low}$ the corresponding weight for the low mixture component, on a logarithmic scale.  Since $f_{high} = 1-f_{low}$, we only show $f_{low}$ here.  The vertical axis shows these same variables, and each panel is the corresponding two-dimensional marginal posteriors for each parameter pair.  The contour region plotted over the sampled posterior represents the 68.3\% credible interval. The one-dimensional histograms are plotted as log density, with the $68.3\%$ credible intervals shown as dotted black lines, and the mode is shown as a vertical solid black line. The two-component Gaussian mixture model characterizes the eccentricity distribution of our sample of planet candidates better than a one-component analysis model.    
}
\label{fig:Nm2_all_corner}
\end{figure*}

\begin{figure*}
\epsscale{0.7}
\plotone{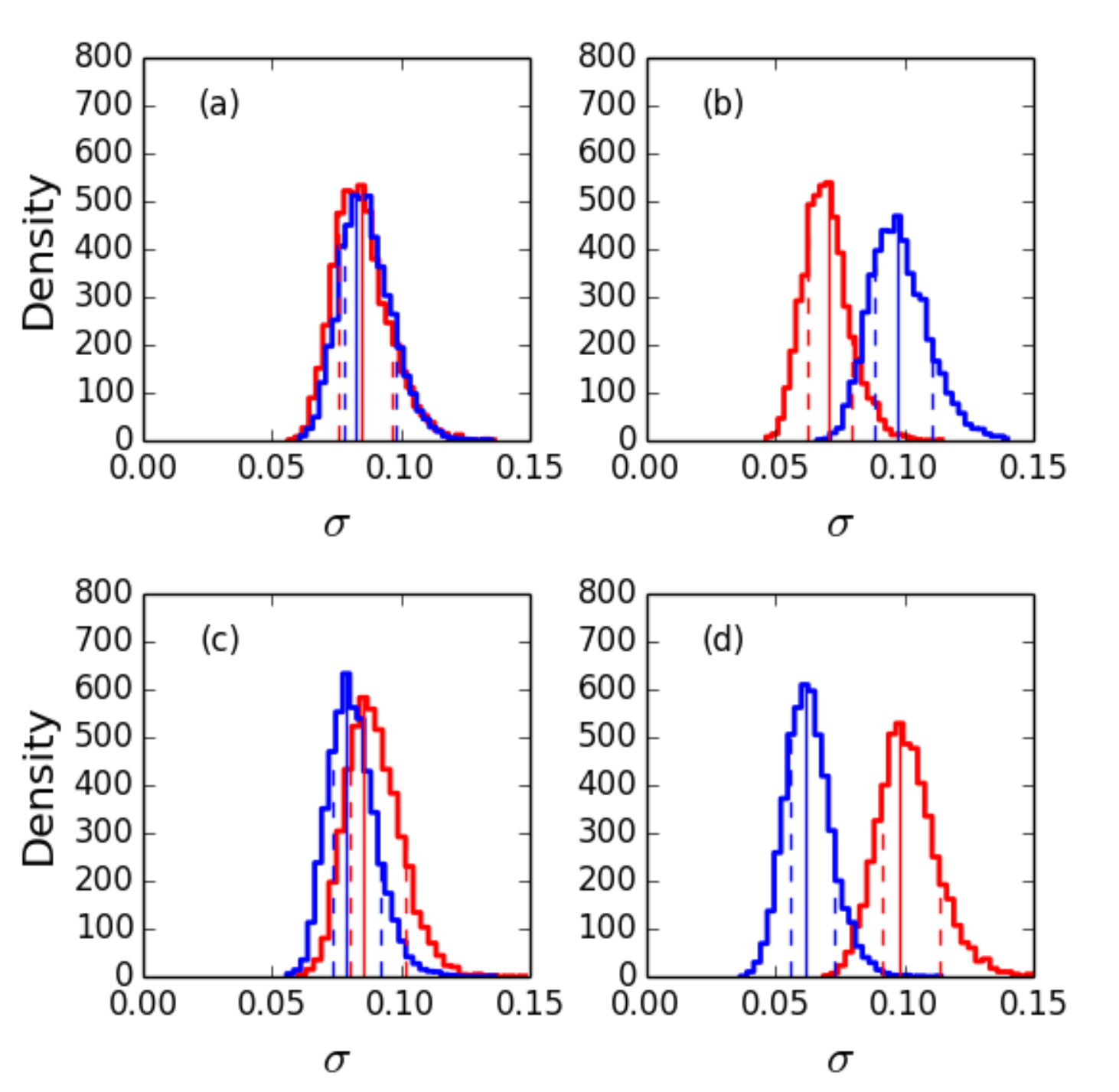}
\caption{One-component Gaussian analysis model applied to subsets of the \textit{Kepler} transit and occultation data.  We apply an HB model to small-valued (blue) and large-valued (red) halves of the \textit{Kepler} short-period planet candidate transit and occultation data, sorted by (a) stellar effective temperature, (b) planet radius, (c) orbital period, and (d) stellar metallicity. The dotted lines correspond to the $68.3\%$ credible intervals and the solid vertical lines correspond to the mode for each posterior distribution. for panel (b) planet radius, and panel (d) stellar metallicity, differences in the small-values and large-valued data subsets merit further investigation.  In order to explore these results further, we analyze these subset using a two-component Gaussian mixture model (see Figure 6). }
\label{fig:Nm1_halves_hk}
\end{figure*}

We thank the entire \textit{Kepler} team for the many years of work that has proven so successful and was critical to this study.  
We acknowledge the SAMSI Bayesian Characterization of Extrasolar Populations working group for discussions that improved this manuscript.
In particular, we thank Merlise Clyde, Darin Ragozzine, Thomas Loredo, David Hogg, Thomas Barclay, and Robert Wolpert for their valuable contributions to the project.  
This work was supported in part by the Pennsylvania State University's Center for Exoplanets and Habitable Worlds, NASA \textit{Kepler} Participating Scientist Program awards NNX12AF73G and NNX14AN76G and NASA Origins of Solar Systems awards NNX13AF61G and NNX14AI76G.   
This material was based upon work partially supported by the National Science Foundation under Grant DMS-1127914 to the Statistical and Applied Mathematical Sciences Institute (SAMSI). Any opinions, findings, and conclusions or recommendations expressed in this material are those of the author(s) and do not necessarily reflect the views of the National Science Foundation.
The authors acknowledge the Institute for CyberScience at The Pennsylvania State University for providing advanced computing resources and services that have contributed to the research results reported in this paper (http://rcc.its.psu.edu), in particular Hoofar Pourzand and William Brouwer.
The authors acknowledge the University of Florida Research Computing (http://researchcomputing.ufl.edu) for providing computational resources and support that have contributed to the research results reported in this publication.
Some/all of the data presented in this paper were obtained from the Mikulski Archive for Space Telescopes (MAST). STScI is operated by the Association of Universities for Research in Astronomy, Inc., under NASA contract NAS5-26555. Support for MAST for non-HST data is provided by the NASA Office of Space Science via grant NNX13AC07G and by other grants and contracts.
This research has made use of the NASA Exoplanet Archive, which is operated by the California Institute of Technology, under contract with the National Aeronautics and Space Administration under the Exoplanet Exploration Program.
We also acknowledge Dan Foreman Mackey's triangle.py open source code for which we adapted for Figure \ref{fig:Nm2_all_corner}.

{\it Facility:} \facility{Kepler}

\bibliographystyle{apj}

\pagebreak

\begin{figure*}
  \epsscale{0.7}
  \plotone{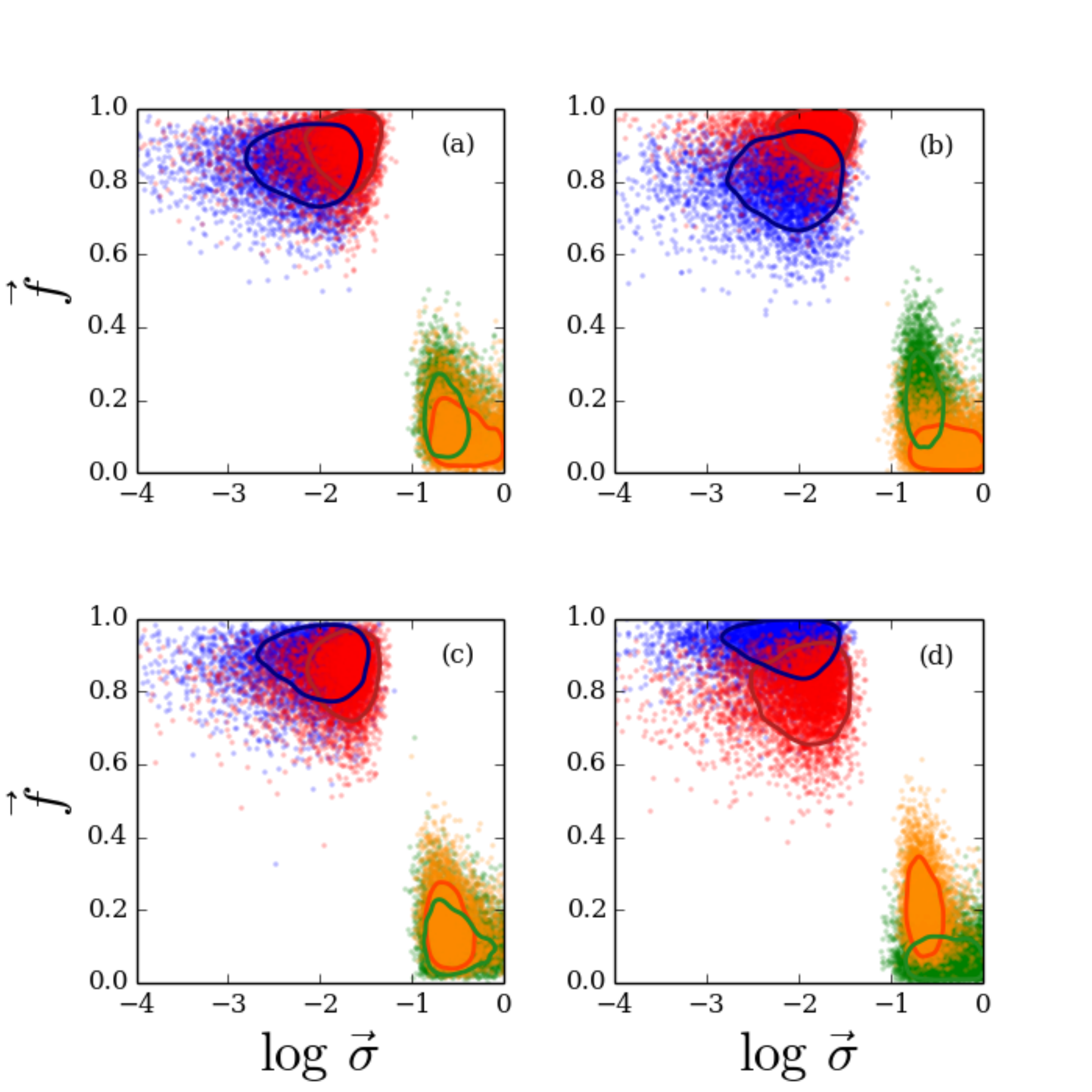}
  \singlespace{
  \caption{Two-component Gaussian mixture model applied to the \textit{Kepler} transit and occultation data.  We apply an HB model to small- and large-valued halves of the short period \textit{Kepler} candidate occultation data, sorted by stellar effective temperature (a),  planet radius (b), orbital period (c), and stellar metallicity (d). The full sample is divided into two equally sized small- and large-value subsets before being processed through our HB model. The small-valued subset of data corresponds to the blue and green clusters, and the large-valued subset corresponds to the red and orange clusters.  The two groups of clusters represent samples of the posterior distribution for the hyperparameter vector, in this case for $\sigma_{low}$ and $f_{low}$ (top left group of clusters), and $\sigma_{high}$ and $f_{high}$ (bottom right group of clusters). 
The data are plotted with the vertical axis representing the low value of the mixture fraction, $f_{low}$, in green and orange, and, $f_{high}$, in blue and red for the two subsets of sorted data shown.  
The contours represent 68.3\% credible intervals.  
Interestingly, for planet radius (b), and stellar metallicity (d), the posteriors of the mixture fractions for the planet candidates with large-valued planet radii and for small-valued host star metallicities are consistent with 0 and 1, indicating only one population is required to accurately model the eccentricity distribution for these subsets of planet candidates. For planet radius and stellar metallicity, we also see that the two-component population models are somewhat different for small and large value subsets.}}
\label{fig:Nm2_halves_hk}
\end{figure*}

\begin{figure*}
  \epsscale{0.5}
  \plotone{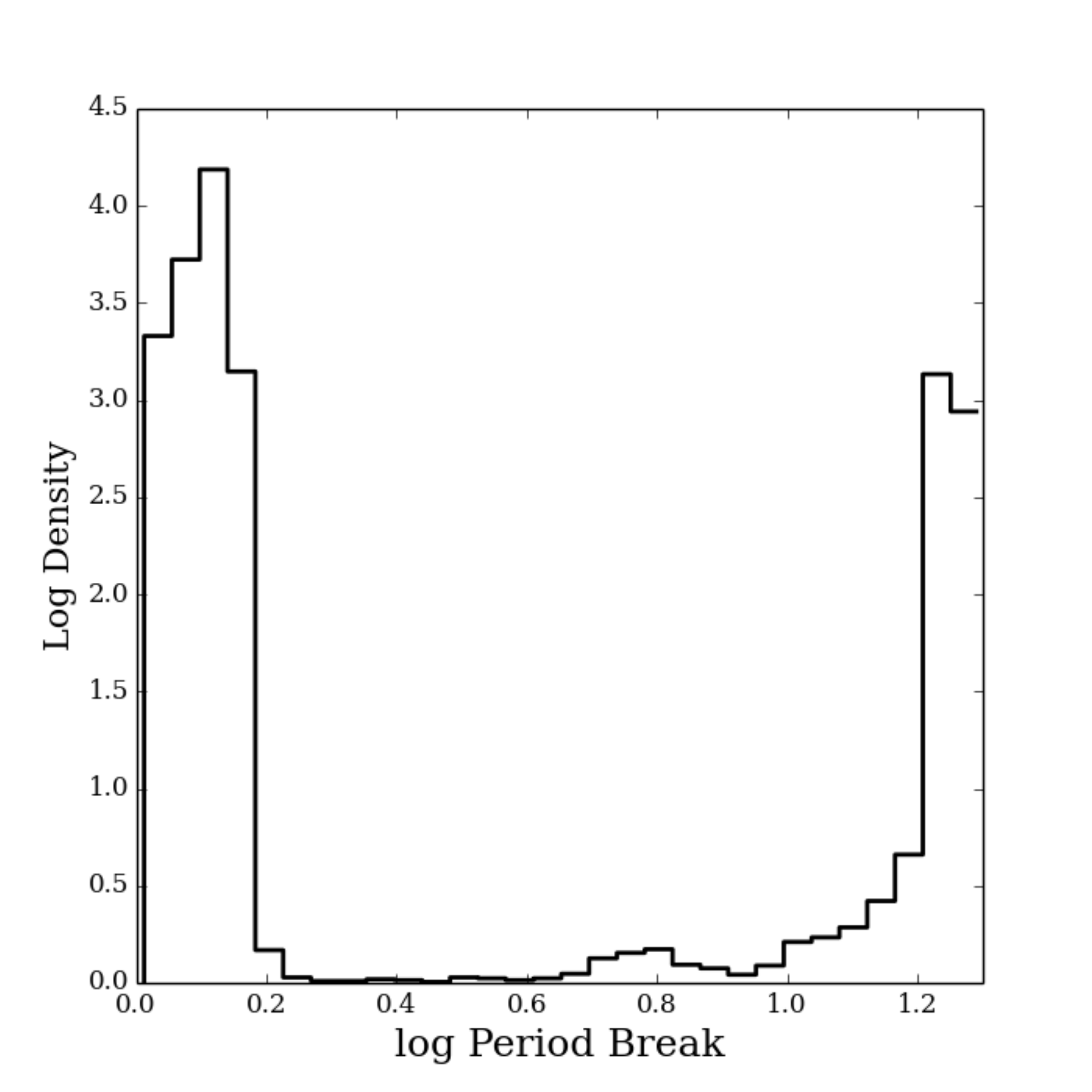}
  \caption{Marginal posterior distribution for the critical period break point from a joint period-eccentricity distribution HB model. We analyze the full dataset using an HB model that allows for the eccentricity distribution to differ depending on whether the orbital period is greater or less than the critical period break point.  We infer that the present data does not allow us to empirically identify a period cutoff that depicts two populations.}
\label{fig:period_break}
\end{figure*}

\begin{deluxetable}{ccccccc}
\tablecaption{Eccentricity Dataset}
\tablenum{1}

\tablehead{\colhead{KOI} & \colhead{$e\cos\omega$} & \colhead{$\sigma_{+e\cos\omega}$} & \colhead{$\sigma_{-e\cos\omega}$} & \colhead{$e\sin\omega$} & \colhead{$ \sigma_{+e\sin\omega}$} & \colhead{$\sigma_{-e\sin\omega}$}  } 

\startdata
$13.01$ & $0.00379$ & $0.00073$ & $0.00073$ & $0.32343$ & $0.01569$ & $0.01559$\\
$17.01$ & $-0.00038$ & $0.03414$ & $0.02824$ & $-0.00144$ & $0.04517$ & $0.04379$\\
$18.01$ & $0.02051$ & $0.03321$ & $0.03515$ & $-0.00965$ & $0.04659$ & $0.05250$\\
$20.01$ & $-0.01868$ & $0.05117$ & $0.04820$ & $0.00010$ & $0.05112$ & $0.05023$\\
$22.01$ & $-0.01123$ & $0.05055$ & $0.05328$ & $-0.00137$ & $0.04962$ & $0.05033$\\
$97.01$ & $-0.00353$ & $0.02024$ & $0.02003$ & $-0.00336$ & $0.03877$ & $0.04284$\\
$98.01$ & $0.00993$ & $0.04816$ & $0.04114$ & $0.00040$ & $0.04973$ & $0.04662$\\
$127.01$ & $0.02353$ & $0.05076$ & $0.05854$ & $0.00120$ & $0.05051$ & $0.05272$\\
$128.01$ & $-0.01801$ & $0.03372$ & $0.03033$ & $0.00044$ & $0.04655$ & $0.04697$\\
$131.01$ & $-0.01178$ & $0.04109$ & $0.04290$ & $-0.00036$ & $0.04971$ & $0.04784$\\
$135.01$ & $-0.04691$ & $0.04427$ & $0.03340$ & $0.00148$ & $0.05046$ & $0.05497$\\
$183.01$ & $0.01357$ & $0.05300$ & $0.04929$ & $0.00246$ & $0.05187$ & $0.05141$\\
$186.01$ & $-0.01337$ & $0.04675$ & $0.04576$ & $0.00049$ & $0.05070$ & $0.04854$\\
$188.01$ & $0.00430$ & $0.03649$ & $0.03731$ & $-0.00252$ & $0.04584$ & $0.04577$\\
$200.01$ & $0.01366$ & $0.05739$ & $0.06048$ & $-0.00142$ & $0.05184$ & $0.05280$\\
$202.01$ & $0.00563$ & $0.04669$ & $0.04344$ & $-0.00377$ & $0.04464$ & $0.04659$\\
$203.01$ & $0.05014$ & $0.02115$ & $0.02894$ & $0.00126$ & $0.05447$ & $0.05276$\\
$204.01$ & $0.01586$ & $0.04639$ & $0.05328$ & $0.00028$ & $0.05031$ & $0.04986$\\
$206.01$ & $-0.01727$ & $0.05781$ & $0.06547$ & $-0.00069$ & $0.05229$ & $0.05331$\\
$254.01$ & $-0.03065$ & $0.05882$ & $0.05747$ & $-0.00141$ & $0.05432$ & $0.05094$\\
$421.01$ & $0.00351$ & $0.05653$ & $0.05293$ & $-0.00169$ & $0.05077$ & $0.05120$\\
$607.01$ & $-0.0027$ & $0.04021$ & $0.03915$ & $-0.00011$ & $0.04710$ & $0.04763$\\
$611.01$ & $0.03344$ & $0.05965$ & $0.05111$ & $0.01742$ & $0.05413$ & $0.05283$\\
$728.01$ & $0.00330$ & $0.04798$ & $0.05457$ & $0.00366$ & $0.05367$ & $0.05392$\\
$760.01$ & $0.01367$ & $0.03489$ & $0.03363$ & $0.00139$ & $0.04686$ & $0.04647$\\
$767.01$ & $-0.00490$ & $0.04486$ & $0.04563$ & $-0.00099$ & $0.05053$ & $0.04772$\\
$774.01$ & $-0.16220$ & $0.00638$ & $0.00332$ & $-0.00802$ & $0.06739$ & $0.05652$\\
$791.01$ & $0.01588$ & $0.02676$ & $0.02860$ & $0.00223$ & $0.04661$ & $0.04651$\\
$797.01$ & $0.04823$ & $0.03688$ & $0.05262$ & $0.01784$ & $0.06601$ & $0.05972$\\
$801.01$ & $0.02502$ & $0.05106$ & $0.04972$ & $-0.00221$ & $0.05348$ & $0.05067$\\
$805.01$ & $0.38761$ & $0.00080$ & $0.00115$ & $0.02390$ & $0.02642$ & $0.02648$\\
$823.01$ & $-0.00629$ & $0.00274$ & $0.00284$ & $-0.35185$ & $0.01225$ & $0.01219$\\
$830.01$ & $0.00997$ & $0.05621$ & $0.05942$ & $-0.00098$ & $0.05115$ & $0.05116$\\
$850.01$ & $-0.01543$ & $0.05830$ & $0.06948$ & $0.00094$ & $0.05208$ & $0.05235$\\
$883.01$ & $-0.02412$ & $0.06865$ & $0.06509$ & $0.00065$ & $0.05498$ & $0.05499$\\
$890.01$ & $0.00621$ & $0.05985$ & $0.03740$ & $0.00096$ & $0.04949$ & $0.04824$\\
$895.01$ & $-0.06154$ & $0.01675$ & $0.01243$ & $-0.00275$ & $0.06048$ & $0.05811$\\
$897.01$ & $0.00571$ & $0.05530$ & $0.04969$ & $0.00083$ & $0.04771$ & $0.05155$\\
$908.01$ & $-0.00760$ & $0.04825$ & $0.04482$ & $-0.00138$ & $0.04992$ & $0.04960$\\
$913.01$ & $0.00433$ & $0.05289$ & $0.04532$ & $0.00276$ & $0.04787$ & $0.05067$\\
$929.01$ & $0.00446$ & $0.03132$ & $0.03626$ & $-2\e{-05}$ & $0.04638$ & $0.04677$\\
$931.01$ & $-0.01950$ & $0.06144$ & $0.06634$ & $0.00042$ & $0.05485$ & $0.05313$\\
$1066.01$ & $-0.05694$ & $0.06242$ & $0.02904$ & $-0.00046$ & $0.05689$ & $0.05470$\\
$1176.01$ & $0.01372$ & $0.04534$ & $0.04789$ & $-0.00070$ & $0.04914$ & $0.04876$\\
$1227.01$ & $0.00424$ & $0.03187$ & $0.04877$ & $-0.30367$ & $0.05812$ & $0.04006$\\
$1391.01$ & $-0.02053$ & $0.02736$ & $0.02283$ & $-3\e{-05}$ & $0.04203$ & $0.04201$\\
$1456.01$ & $0.00524$ & $0.03596$ & $0.03497$ & $-0.00229$ & $0.04655$ & $0.04747$\\
$1457.01$ & $-0.00701$ & $0.04232$ & $0.02725$ & $-0.00038$ & $0.04791$ & $0.04625$\\
$1781.01$ & $0.07127$ & $0.01272$ & $0.02821$ & $0.00197$ & $0.05888$ & $0.05369$\\
$1793.01$ & $0.00685$ & $0.04855$ & $0.04713$ & $0.00578$ & $0.04827$ & $0.04871$\\
\enddata
\tablecomments{Result for $e \cos \omega$ and $e \sin \omega$ from MCMC.  See \textsection 2.1 for details on how these values are calculated.}


\label{table1:data}

\end{deluxetable}

\begin{deluxetable}{cccc}
\tablecaption{Model Datasets}
\tablenum{2}

\tablehead{\colhead{Model Name} & \colhead{Np} & \colhead{$\sigma_h$} & \colhead{$\sigma_k$}  } 

\startdata
half & $25$ & $0.040$ & $0.080$ \\
good & $50$ & $0.040$ & $0.080$ \\
better & $50$ & $0.001$ & $0.001$ \\
best & $500$ & $0.001$ & $0.001$ \\
\enddata


\tablecomments{Values indicating the quantity and quality of the suite of simulated observations used in our analysis. Datasets labeled ``good" (``half") consist of $50$ ($25$) planets with measurement uncertainties of $0.04$ and $0.08$ for $h$ and $k$ respectively. These datasets are designed to be similar to our actual transit and occultation dataset for both $h$ and $k$.  Datasets labeled ``better" (``best") contain $50$ ($500$) planets with measurement uncertainties of $0.001$, and are designed to forecast the power of this method and model setup when the quantity of real data grows and the quality of data is improved upon (better measurement uncertainty).}

\label{table2:models}
\end{deluxetable}

\begin{deluxetable}{cccc}
\tablecaption{Median K-S Statistic for $h$}
\tablenum{3}

\tablehead{\colhead{} & \colhead{} & \colhead{Analysis Model} & \colhead{} \\ 
\colhead{Generalized Model Name} & \colhead{($N_{m}=1$)} & \colhead{($N_{m}=2$)} & \colhead{($N_{m}=3$)} } 

\startdata
R1 half & $0.1310$ & $0.1615$ & $0.1740$\\
R1 good & $0.1550$ & $0.1650$ & $0.1700$\\
R1 better & $0.0990$ & $0.1150$ & $0.1275$\\
R1 best & $0.0400$ & $0.0500$ & $0.0550$\\
R2 half & $0.2365$ & $0.2100$ & $0.2715$\\
R2 good & $0.2545$ & $0.1875$ & $0.2330$\\
R2 better & $0.1890$ & $0.0825$ & $0.1590$\\
R2 best & $0.1705$ & $0.0660$ & $0.1485$\\
R3 half & $0.1535$ & $0.1215$ & $0.1415$\\
R3 good & $0.2560$ & $0.2085$ & $0.2285$\\
R3 better & $0.1270$ & $0.1035$ & $0.1350$\\
R3 best & $0.1726$ & $0.0835$ & $0.2142$\\

\enddata

\tablecomments{Results of validation and sensitivity analysis of three hierarchical models for eccentricities.  
Shown in this table are Kolmogorov-Smirnov (K-S) statistics comparing datasets of simulated observations with datasets generated using the posteriors of the hyperparameters from an HB model that analyzed the same set of simulated observations (comparing input to output to test model). Here R1, R2, and R3 represent a one-, two- and three-component Gaussian mixture model, respectively. Table \ref{table2:models} summarizes the different quantity and quality of simulated observations used in this analysis.  A two-component Gaussian mixture model does well across the majority of simulated datasets. See \textsection 4.2.2 for a detailed interpretation of these results.}

\label{table3:KSstats}
\end{deluxetable}

\end{document}